\documentclass[usenatbib,usegraphicx,usedcolumn]{mn2e}
\usepackage{times,amssymb,amsmath,mathrsfs}
\usepackage{relsize}
\usepackage[usenames,dvipsnames]{color}

\usepackage[backref,breaklinks,colorlinks,citecolor=blue]{hyperref}
\usepackage{rotating}

\title[Moving mesh unstaggered constrained transport]{A moving mesh unstaggered constrained transport scheme for magnetohydrodynamics}
\author[P. Mocz et. al.]{Philip Mocz$^{1}$\thanks{E-mail: pmocz@cfa.harvard.edu (PM)}, 
R\"udiger Pakmor$^{3}$,
Volker Springel$^{3,4}$,
Mark Vogelsberger$^{2}$,
\newauthor
Federico Marinacci$^{2}$,
Lars Hernquist$^{1}$ \\
$^{1}$Harvard-Smithsonian Center for Astrophysics, 60 Garden Street, Cambridge, MA 02138, USA \\
$^{2}$Kavli Institute for Astrophysics and Space Research, Massachusetts Institute of Technology, Cambridge, MA 02139, USA\\
$^{3}$Heidelberger Institut f\"ur Theoretische Studien, Schloss-Wolfsbrunnenweg 35, 69118 Heidelberg, Germany\\
$^{4}$Zentrum f\"ur Astronomie der Universit\"at Heidelberg, Astronomisches
Recheninstitut, M\"onchhofstr. 12-14, 69120 Heidelberg, Germany
}
\begin{document}

\date{submitted to MNRAS, June 2016}

\pagerange{\pageref{firstpage}--\pageref{lastpage}} \pubyear{2013}

\maketitle

\label{firstpage}
%###########################################################################################################

\begin{abstract} We present a constrained transport (CT) algorithm for solving the 3D ideal magnetohydrodynamic (MHD) equations on a moving mesh, which maintains the divergence-free condition on the magnetic field to machine-precision.  Our CT scheme uses an unstructured representation of the magnetic vector potential, making the numerical method simple and computationally efficient. The scheme is implemented in the moving mesh code \textsc{Arepo}. We demonstrate the performance of the approach with simulations of driven MHD turbulence, a magnetized disc galaxy, and a cosmological volume with primordial magnetic field. We compare the outcomes of these experiments to those obtained with a previously implemented Powell divergence-cleaning scheme.  While CT and the Powell technique yield similar results in idealized test problems, some differences are seen in situations more representative of astrophysical flows. In the turbulence simulations, the Powell cleaning scheme artificially grows the mean magnetic field, while CT maintains this conserved quantity of ideal MHD. In the disc simulation, CT gives slower magnetic field growth rate and saturates to equipartition between the turbulent kinetic energy and magnetic energy, whereas Powell cleaning produces a dynamically dominant magnetic field. Such difference has been observed in adaptive-mesh refinement codes with CT and smoothed-particle hydrodynamics codes with divergence-cleaning. In the cosmological simulation, both approaches give similar magnetic amplification, but Powell exhibits more cell-level noise. CT methods in general are more accurate than divergence-cleaning techniques, and, when coupled to a moving mesh can exploit the advantages of automatic spatial/temporal adaptivity and reduced advection errors, allowing for improved astrophysical MHD simulations. 
\end{abstract}


\section{Introduction}\label{sec:intro}

The need for performing accurate, spatially- and temporally-adaptive 
magnetohydrodynamic (MHD) simulations in astrophysics is evident. 
Magnetic fields are prevalent in a range of astrophysical systems from 
cosmological \citep{2015MNRAS.453.3999M}, cluster \citep{2012MNRAS.419.3319M}, and galaxy \citep{2009ApJ...696...96W,2010A&A...523A..72D,2012MNRAS.422.2152B,2014ApJ...783L..20P,2016MNRAS.457.1722R} scales, to the turbulent interstellar 
medium \citep{2012ApJ...750...13C,2012ApJ...761..156F,2013ApJ...766...97M,2015MNRAS.450.4035F}, accretion around black holes \citep{2014MNRAS.441.3177M,2014MNRAS.439..503S}, 
mergers of compact objects \citep{2015ApJ...806L...1Z}, tidal 
disruption events \citep{2014MNRAS.445.3919K}, and others.  Often, these systems exhibit a large 
dynamic range in physical and temporal scales.  For example, a simulation 
box may contain significant regions of low density gas combined with 
concentrated volumes where most of the material is present and 
dynamical time-scales are the shortest. Adaptive, minimally-diffusive 
schemes are necessary for simulating such configurations precisely, 
given the limitations in memory and computation speed of modern 
supercomputing technology.

Solving the ideal MHD equations is more challenging numerically than 
evolving the inviscid, magnetic-free Euler equations, due to the 
divergence free nature of the magnetic field 
($\nabla\cdot\mathbf{B}=0$) from Maxwell's equations.  This condition 
needs to be maintained in the discrete representation of the fluid in 
order for a numerical solver to be both stable and accurate. The 
finite-volume method is a standard approach for solving the Euler 
equations on a mesh.  However, this technique fails for the MHD 
equations because divergence errors can cause the magnetic fields to 
blow up.  On static, regular, Cartesian meshes, the state-of-the art 
solution to this problem is to use the constrained transport (CT) 
scheme \citep{1988ApJ...332..659E}, which maintains the discretized 
divergence of $\mathbf{B}$ to zero exactly at machine precision. This 
method originally comes from the staggered-mesh method of 
\cite{1966ITAP...14..302Y} for electromagnetism in a vacuum. 
Conceptually, the $\nabla\cdot\mathbf{B}=0$ constraint is exactly 
maintained with CT by representing the magnetic field as cell 
face-averaged quantities (as opposed to the usual choice of 
volume-averages), and making use of Stokes' Theorem.

The CT method does not generalize easily, however, to moving meshes 
(e.g. \citealt{2010MNRAS.401..791S}, \citealt{2011ApJS..197...15D}, 
\citealt{2012ApJ...758..103G}), and may not be at all applicable to 
meshless Lagrangian approaches (e.g. smoothed particle hydrodynamics (SPH)
\citealt{2015arXiv150504494T}, or the volume `overlap' method of 
\citealt{2015MNRAS.450...53H,2016MNRAS.455...51H}).  Consequently, 
numerical solvers in a Lagrangian setting have resorted to magnetic 
field cleaning schemes, such of those of \cite{2002JCoPh.175..645D} and 
\cite{1999JCoPh.154..284P}, which, while stable, may have unwanted 
numerical side-effects. This has been an unfortunate situation for 
these Lagrangian approaches, because otherwise they offer many 
advantages over static mesh codes, including automatic spatial 
adaptability, significantly larger CFL-limited timesteps for high Mach 
number flows, and reduced advection errors.

In this paper, we overcome this limitation by developing and presenting 
an unstaggered CT method for solving the 3D ideal MHD equations on 
moving meshes. Seminal ideas for our new technique come from our recent 
paper \cite{2014MNRAS.442...43M}, where we generalized the staggered CT 
approach to a 2D moving mesh. However, we have now modified and refined 
the ideas further to improve the efficiency and simplicity of the 
algorithm.  The new scheme employs an unstructured (cell-centred) 
formulation, using the magnetic vector potential, which makes it 
straightforward to adapt to moving mesh codes.  We have implemented 
this new unstaggered CT method into \textsc{Arepo} 
\citep{2010MNRAS.401..791S}, which is a state-of-the-art moving mesh 
code for astrophysical flows.

We provide a brief history of the development of moving mesh MHD 
solvers for astrophysics, to place our work in context.  The 
finite-volume moving mesh formulation for the Euler equations was 
developed for astrophysical simulations by \cite{2010MNRAS.401..791S}. 
The MHD equations were first solved on a moving mesh with some success 
in \cite{2011ApJS..197...15D,2011MNRAS.418.1392P,2012ApJ...758..103G}, 
using the Dedner hyperbolic divergence-cleaning scheme. However, 
numerical limitations of these approaches have been reported. 
\cite{2013MNRAS.432..176P} implemented a Powell cleaning technique for 
a moving mesh, which showed improved stability (even in very dynamic 
environments) but larger divergence errors (but still small enough in 
many cases to not affect dynamics). The Powell scheme has been used for 
science applications, including the study of magnetic fields in disc 
galaxies \citep{2014ApJ...783L..20P}, Carbon-Oxygen white dwarf mergers 
\citep{2015ApJ...806L...1Z}, and the large-scale properties of 
simulated cosmic magnetic fields and effects on the galaxy population \citep{2015MNRAS.453.3999M,2016MNRAS.456L..69M}. 
\cite{2014MNRAS.442...43M} extended the standard CT algorithm to a 2D 
moving Voronoi mesh, keeping track of face-averaged magnetic fields and 
remapping onto new faces that appear as the mesh changes connectivity. 
We have simplified this idea in the present paper to easily and 
efficiently extend it to a 3D moving mesh code.

The paper is organized as follows. We present the unstaggered moving 
mesh CT scheme in Section~\ref{sec:methods}. In Section~\ref{sec:tests} 
we show the results of test problems (Orszag-Tang, circularly polarized Alfv\'en wave, turbulent box, magnetic disc, 
cosmological volume) and compare with Powell cleaning. We discuss the 
advantages of the new algorithm and 
offer concluding remarks in Section~\ref{sec:discussion}.

\section{Numerical Method}\label{sec:methods}

In this section we describe our numerical method for an unstructured CT 
solver on a moving Voronoi mesh. The ideal MHD equation and some 
notation are presented in Section~\ref{sec:mhd}. The CT algorithm is 
detailed in Section~\ref{sec:ct}.  A procedure for implementing the 
approach in a periodic domain are discussed in Section~\ref{sec:A}. 

\subsection{The magnetohydrodynamic equations}\label{sec:mhd}

The ideal MHD equations are represented in conservation law form by:
\begin{equation}
\frac{\partial \mathbf{U}}{\partial t} + \nabla \cdot \mathbf{F} = 0
\end{equation}
where $\mathbf{U}$ is the vector of the conserved variables and $\mathbf{F}(\mathbf{U})$ is the flux:
\begin{equation}
\mathbf{U} = \begin{pmatrix} \rho \\ \rho\mathbf{v} \\ \rho e \\ \mathbf{B} \end{pmatrix},
\,\,\,\,\,\,
\mathbf{F}(\mathbf{U}) = 
\begin{pmatrix} \rho\mathbf{v} \\ \rho\mathbf{v}\mathbf{v}^T + p -\mathbf{B}\mathbf{B}^T \\ \rho e \mathbf{v} + p\mathbf{v} -\mathbf{B}(\mathbf{v}\cdot \mathbf{B}) \\ \mathbf{B}\mathbf{v}^T-\mathbf{v}\mathbf{B}^T \end{pmatrix}
\end{equation}
and $p=p_{\rm gas}+\frac{1}{2}\mathbf{B}^2$ is the total gas pressure,
$e=u+\frac{1}{2}\mathbf{v}^2+\frac{1}{2\rho}\mathbf{B}^2$ is the total
energy per unit mass, and $u$ is the thermal energy per unit mass. The
equation of state for the fluid is given by the ideal gas law
$p=(\gamma-1)\rho u$.

The mass density ($\rho$), momentum density ($\rho\mathbf{v}$), and 
energy density ($\rho e$) are evolved according to the second-order, 
Runge-Kutta time integrator, finite volume approach presented in 
\cite{2016MNRAS.455.1134P}.  However, the magnetic field is a special 
quantity because of the divergence-free constraint 
$\nabla\cdot\mathbf{B} = 0$. We now describe how to evolve this 
quantity, maintaining the divergence-free condition to machine 
precision.

\subsection{Unstaggered constrained transport on a moving mesh}\label{sec:ct}

In the original staggered CT approach, magnetic fields are represented
as fluxes normal to a cell-face, and are updated according to the
induction equation by calculating the contribution from the
electromotive force (EMF) in a loop around the edges of a face. In
this formulation, the net change in the outward normal fluxes through
any closed surface in the domain is zero, which is to say that the
field is maintained divergence-free by Stokes' Theorem.  The
face-averaged representation of the magnetic field has been employed
in \cite{2014MNRAS.442...43M} for a 2D moving mesh.  Here, however, we
reformulate the method in an unstaggered manner to improve simplicity
and efficiency.  In a sense, the two methods are still very similar:
the same information encoded in a staggered CT method can be encoded
in an unstaggered CT scheme.  Namely, the same EMF update terms added
to the magnetic flux through a face can instead be added to the
magnetic vector potential component projected along the sides of the
face.

Consider the magnetic field $\mathbf{B}$ written in terms of the
vector potential $\mathbf{A}$ under the Weyl gauge\footnote{we discuss the possibility of using other gauge choices in Appendix~\ref{appendix}}:
\begin{equation}
\mathbf{B} = \nabla\times\mathbf{A}.
\end{equation}
The vector potential evolves according to the induction equation
\begin{equation}
\frac{\partial \mathbf{A}}{\partial t} = -\mathbf{E}
\end{equation}
where $\mathbf{E}=-\mathbf{v}\times\mathbf{B}$ is the electric field for an ideal MHD fluid.

In our representation, each cell $i$ maintains and evolves the volume integral of the vector potential $\mathbf{Q}_i \equiv \int_{V_i}\mathbf{A}\,dV$. The cell-averaged value $\mathbf{A}_i$ for the vector potential is thus $\mathbf{A}_i \equiv \mathbf{Q}_i / V_i$, where $V_i$ is the cell volume. On a moving mesh, $\mathbf{Q}$ evolves as
\begin{equation}
\frac{d\mathbf{Q}_i}{dt} = -\int_{V_i}\mathbf{E}\,dV - \int_{\partial V_i} -\mathbf{A}\mathbf{w}^T\,d\mathbf{n}
\end{equation}
where $\mathbf{n}$ is the outward normal vector of the cell surface,
and $\mathbf{w}$ is the velocity at which each point in the boundary
of the cell moves. The second integral term is just the advection
due to the mesh motion. That is, the evolution for $\mathbf{Q}_i$ has
a source term due to the electric field, and a flux term due to mesh
motion. The source term is treated in a Strang-split fashion by
applying two half-timesteps before and after evolving the homogeneous
system by one step, similar to the treatment of the gravitational
source terms in \textsc{Arepo}. The flux term due to mesh motion is
treated by taking the upwind value of the magnetic vector potential to
calculate the flux.

In discretized terms, $\mathbf{Q}_i$ is updated in a second-order
approach from time step $n$ to $n+1$ with a Huen's method Runge-Kutta
integrator \citep{2016MNRAS.455.1134P} as
\begin{eqnarray}
\mathbf{Q}^{(n+1)}_i =& \mathbf{Q}^{(n)}_i - \frac{\Delta t}{2} \left( \mathbf{E}^{(n)}_iV^{(n)}_i+   \mathbf{E}^{'}_iV^{'}_i\right) \\
& -  \frac{\Delta t}{2}  \left( 
\sum_j A_{ij}^{(n)}\mathbf{F}^{n}_{ij}(\mathbf{A}^n)
+
\sum_j A_{ij}^{(n+1)}\mathbf{F}^{'}_{ij}(\mathbf{A}^{'})
 \right)
\end{eqnarray}
where $\Delta t$ is the time step, the sum is taken over all
neighbours $j$, $A_{ij}$ is the area of the shared face between cells
$i$ and $j$, $\mathbf{F}_{ij}$ is the upwind flux of $\mathbf{A}$ due
to the mesh motion, and the primed superscript represents the variables
time-extrapolated from time level $(n)$ forward by $\Delta t$. The
flux is evaluated by extrapolating the cell-averaged quantities out to
the faces and taking the upwind value in the Riemann problem. This
update step is symmetric in the sense that it uses the mesh geometry
at both the beginning and end of the timestep to evolve the fluid
quantities.
Note that using the electric fields predicted at cell-centers obtained from evolving the electric field with properly upwinded, shock-capturing Riemann fluxes ($\mathbf{E}^{'}_i$) is reminiscent of the field-interpolated central difference version of constrained transport of \cite{2000JCoPh.161..605T}.

The key, next, is to have a CT mapping that transforms the vector
potentials, $\mathbf{A}_i$, to magnetic fields $\mathbf{B}_i$, while
maintaining a particular discretization of
$\nabla\cdot\mathbf{B}=0$. This value of $\mathbf{B}_i$, which we
denote as $\mathbf{B}_{\rm CT,i}$, overwrites the value that would
have been obtained from the Riemann solver in a finite-volume scheme,
$\mathbf{B}_{\rm FV,i}$. This `corrector' step is the standard CT approach. Sometimes,
optionally, an energy correction step is applied to each cell as well
to maintain consistency of the magnetic field in the total energy
evolution of the fluid at the cost of machine-precision total energy
conservation \citep{1999JCoPh.149..270B},
\begin{equation}
\mathbf{E}_i \leftarrow \mathbf{E}_i + \frac{1}{2}B^2_{\rm CT,i} - \frac{1}{2}B^2_{\rm FV,i}
\end{equation}
although we do not find this necessary for our simulations.

\begin{figure}
\centering
\includegraphics[width=0.47\textwidth]{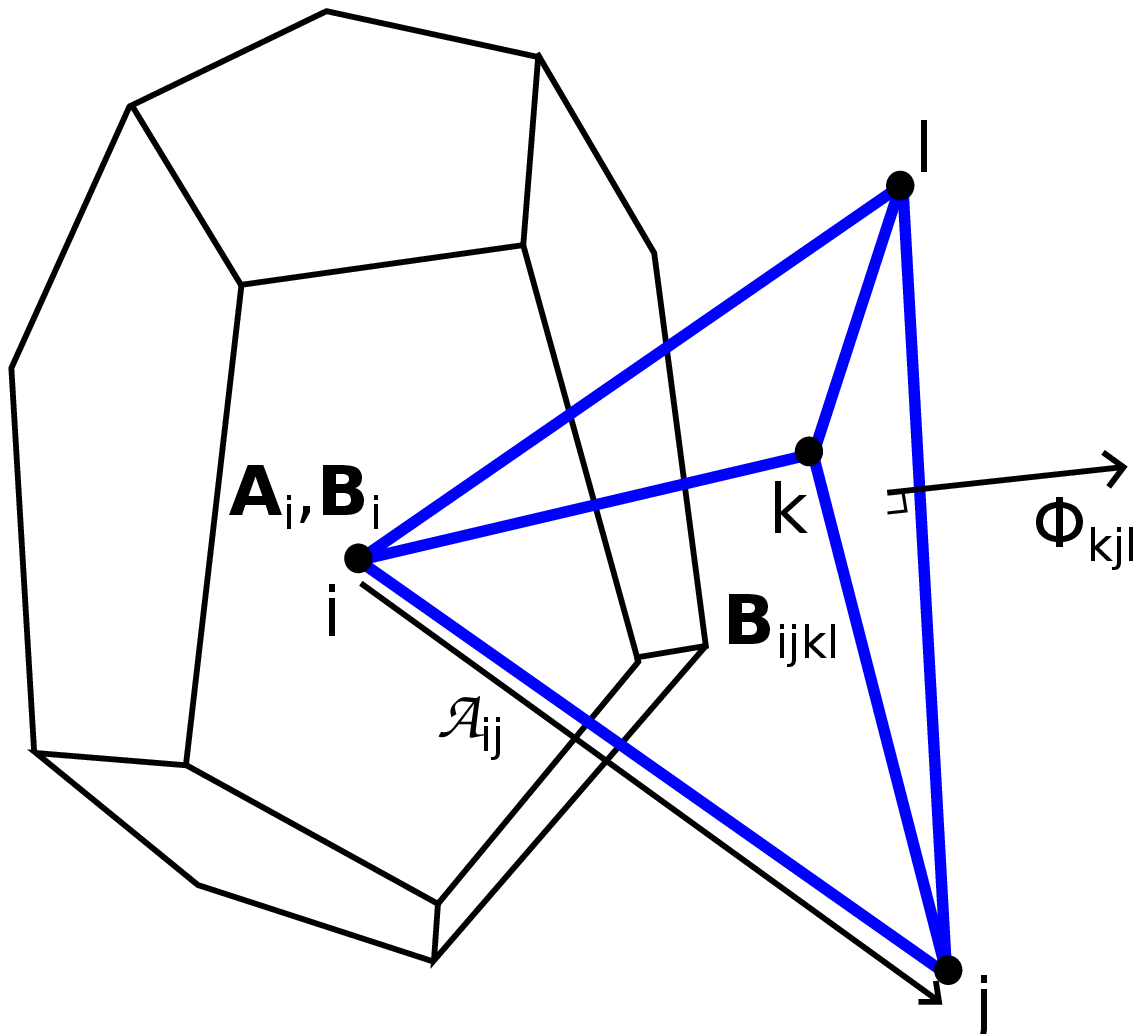} \\
\bigskip
{\large
$\mathbf{A}^{(n)}_i \rightarrow 
\mathcal{A}_{ij} \rightarrow
\Phi_{ijk} \rightarrow
\mathbf{B}_{ijkl} \rightarrow
\mathbf{B}^{(n)}_{i} \rightarrow
\mathbf{A}^{(n+1)}_i$
}
\caption{A geometric representation of the magnetic field quantities
used for the unstaggered CT representation on an unstructured
mesh. The Voronoi cell $i$ is represented by the polyhedron, and one
of the touching Delaunay tetrahedra is shown in blue. Points $i$,
$j$, $k$, and $l$ are mesh generating points. Our method evolves the
volume-averaged magnetic vector potential, which is projected along
Delaunay edges, and used to recover the Delaunay face-averaged
magnetic fluxes, Delaunay volume-averaged magnetic fields, and
consequently cell-centred magnetic fields. The arrowed diagram shows
the order of steps taken to recover the magnetic fields and evolve
the system to the next time step.}
\label{fig:scheme}
\end{figure}

\begin{figure*}
\centering
\begin{tabular}{cccc}
 & \large{moving CT} & \large{static CT} & \large{moving Powell} \\
\rotatebox{90}{\hspace{9 mm} $B$ \,\,\,\,\,  $t=2.0$ \,\,\,\,\, $N=16^3$} &
\includegraphics[width=0.29\textwidth]{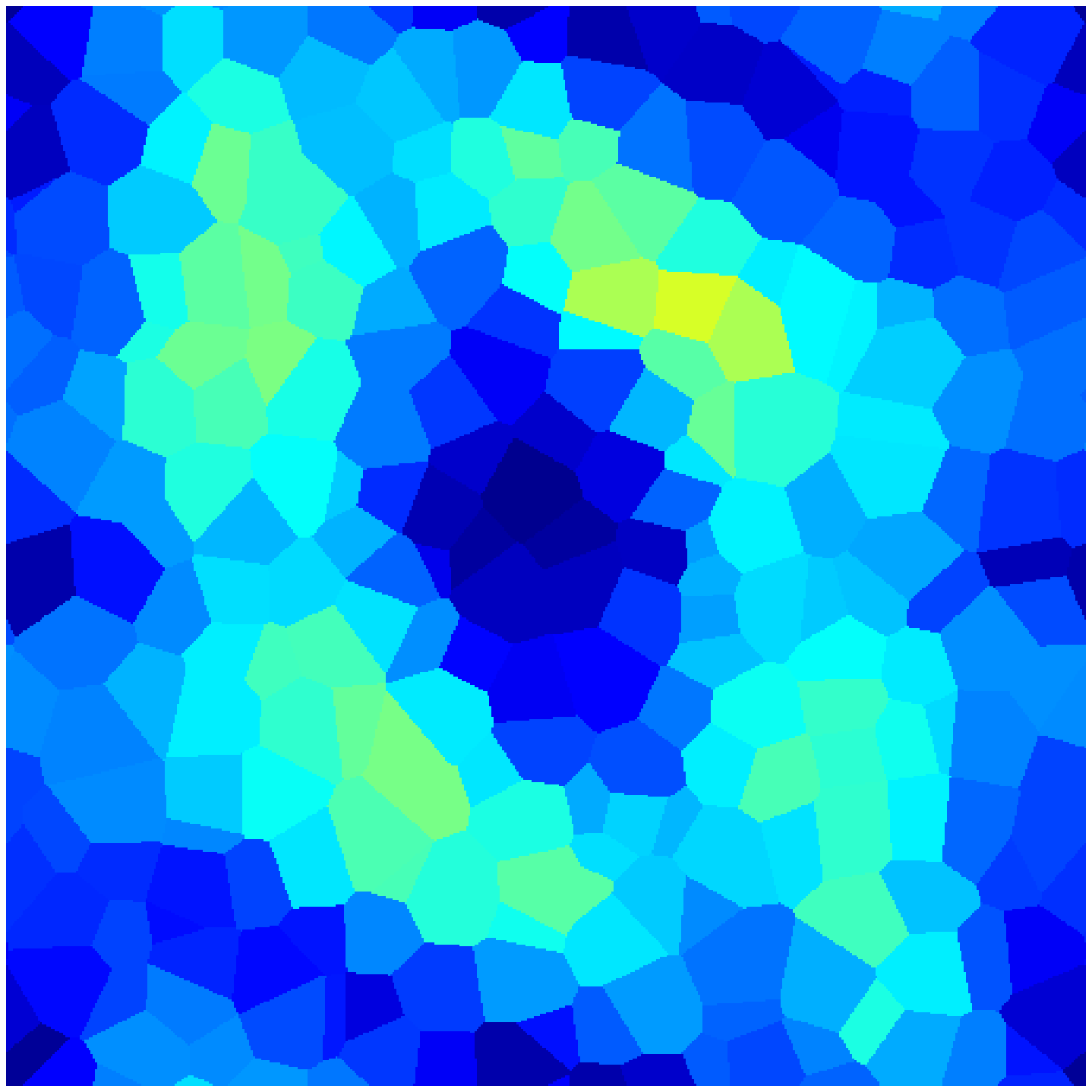} &
\includegraphics[width=0.29\textwidth]{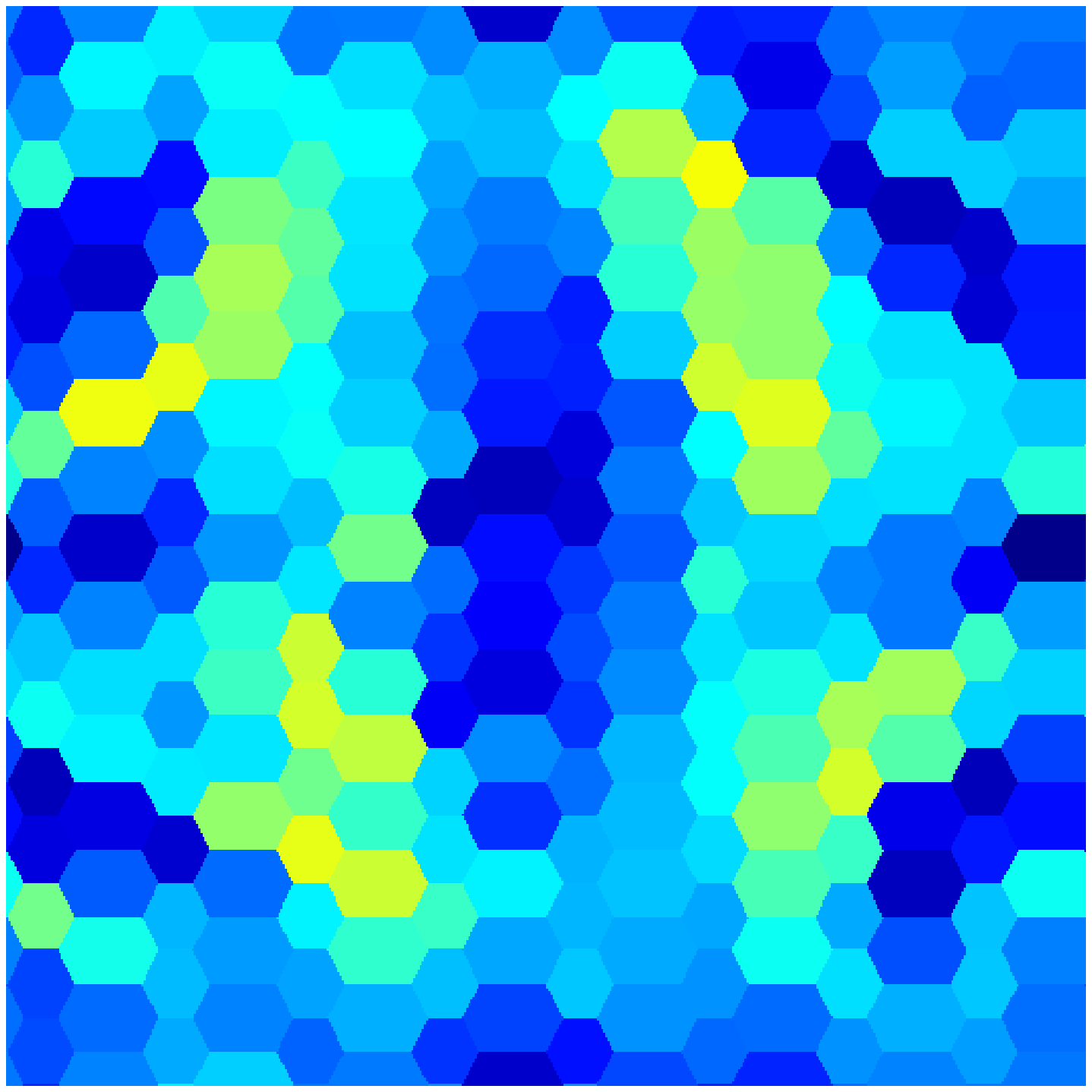} &
\includegraphics[width=0.29\textwidth]{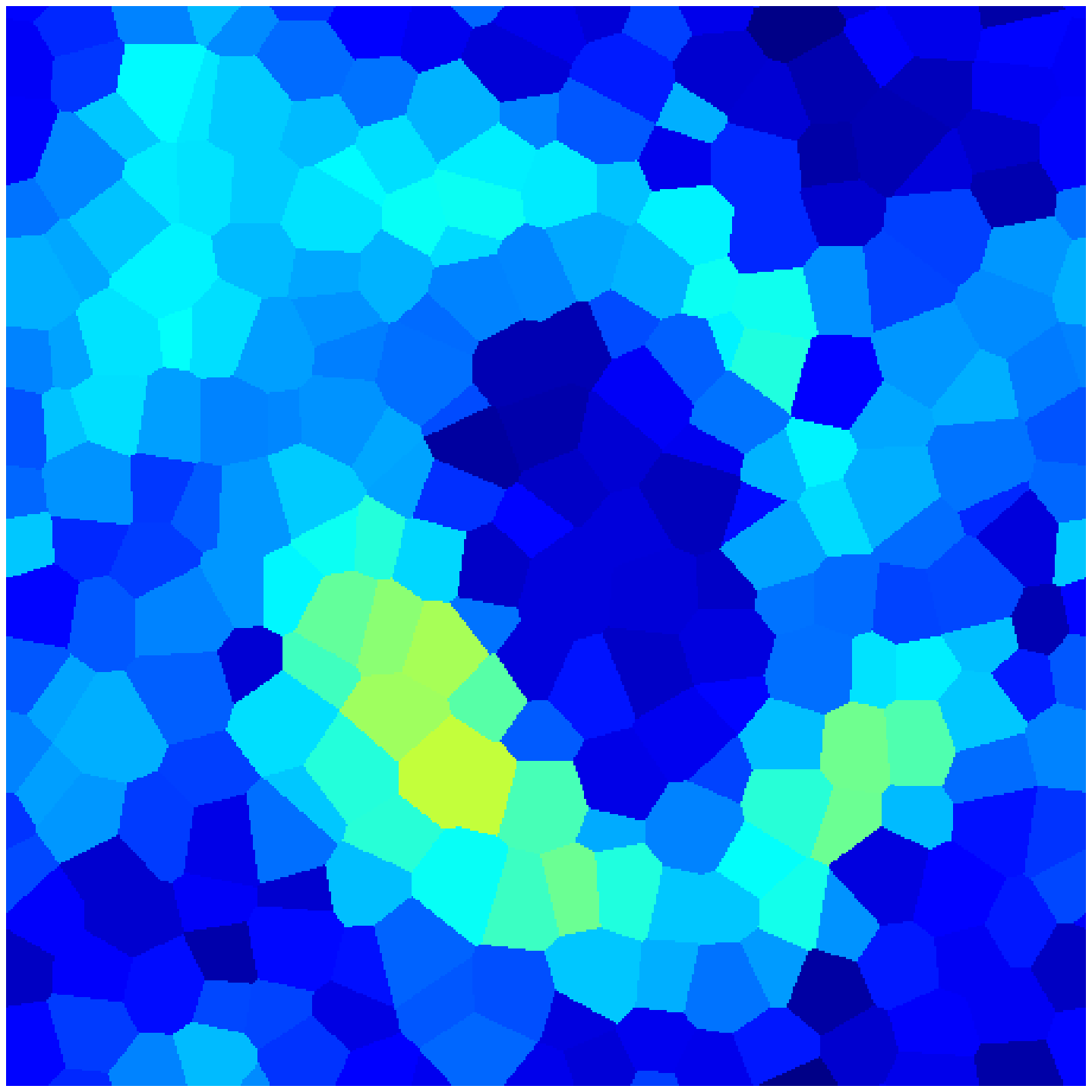} \\
\rotatebox{90}{\hspace{9 mm} $B$ \,\,\,\,\,  $t=2.0$ \,\,\,\,\, $N=64^3$} &
\includegraphics[width=0.29\textwidth]{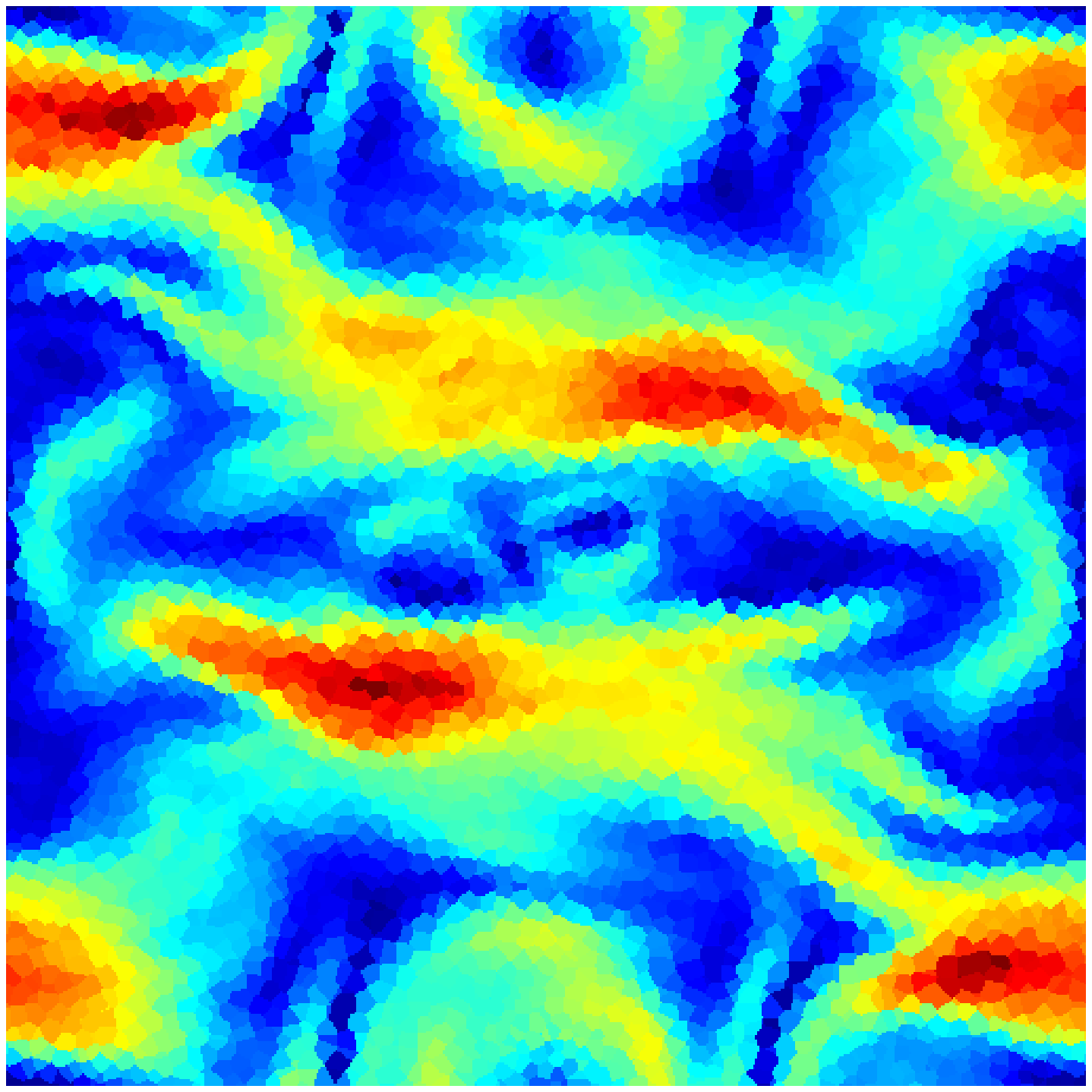} &
\includegraphics[width=0.29\textwidth]{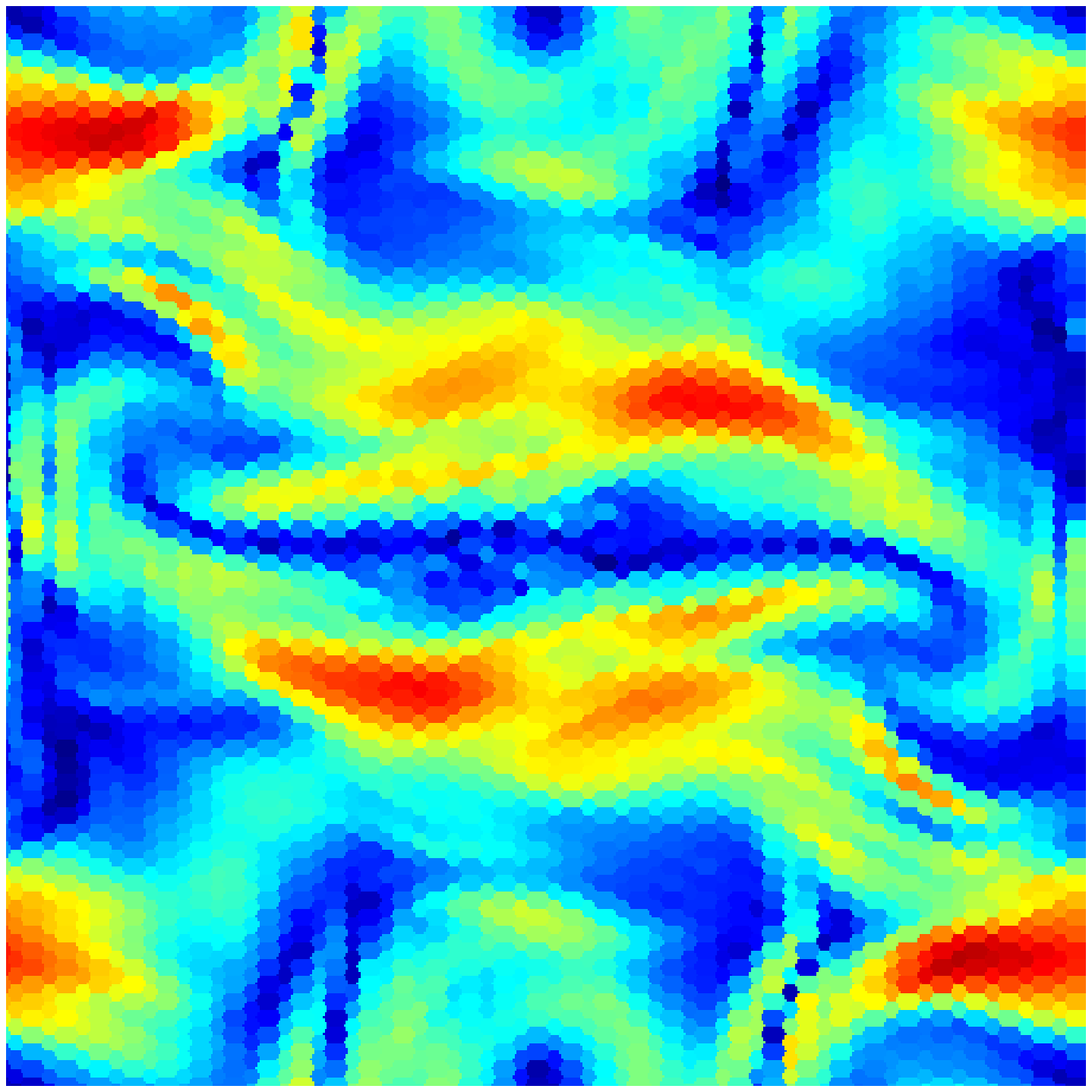} &
\includegraphics[width=0.29\textwidth]{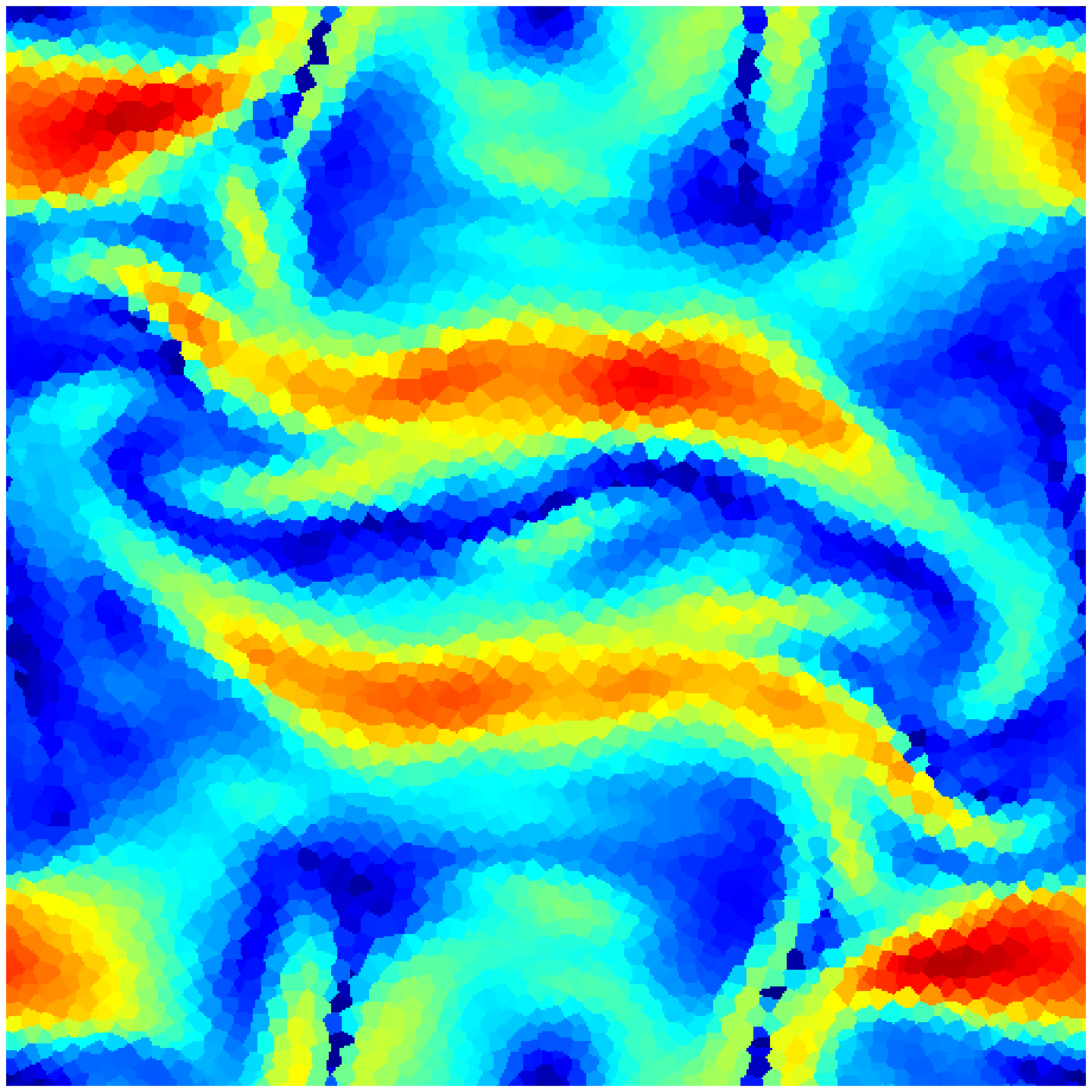} \\
  &
\multicolumn{3}{c}{\includegraphics[width=0.9\textwidth]{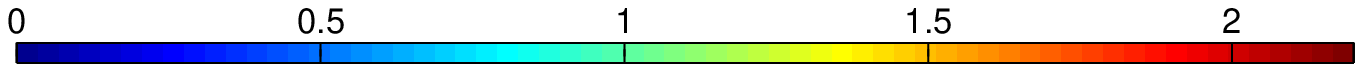}}
\end{tabular}
\caption{A comparison of moving CT, static CT, and moving Powell
schemes for the Orszag-Tang test, as labeled at
the top. The magnetic field strength is
shown at $t=2.0$, simulated at low (upper panels)
and high (lower panels) resolutions. All approaches
give results consistent and accurate results.}
\label{fig:otmain}
\end{figure*}

\begin{figure*}
\centering
\begin{tabular}{ccc|cc}
&  \multicolumn{2}{c}{\textbf{moving CT}} & \multicolumn{2}{c}{\textbf{moving Powell}}  \\
& \includegraphics[width=0.22\textwidth]{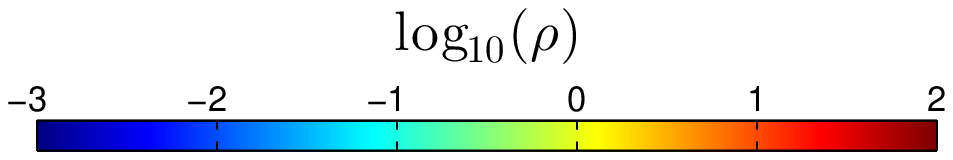}
& \includegraphics[width=0.22\textwidth]{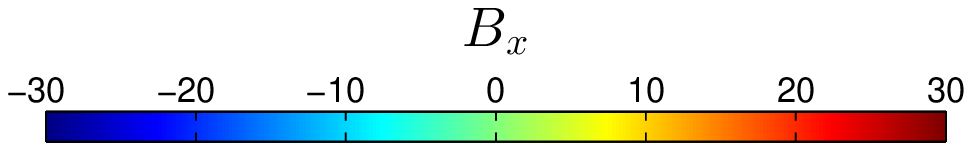}
& \includegraphics[width=0.22\textwidth]{turbcbrho.eps}
& \includegraphics[width=0.22\textwidth]{turbcbbx.eps} \\
\begin{turn}{90} \qquad\qquad $t=2\tau_{\rm eddy}$ \end{turn}
&
\includegraphics[width=0.22\textwidth]{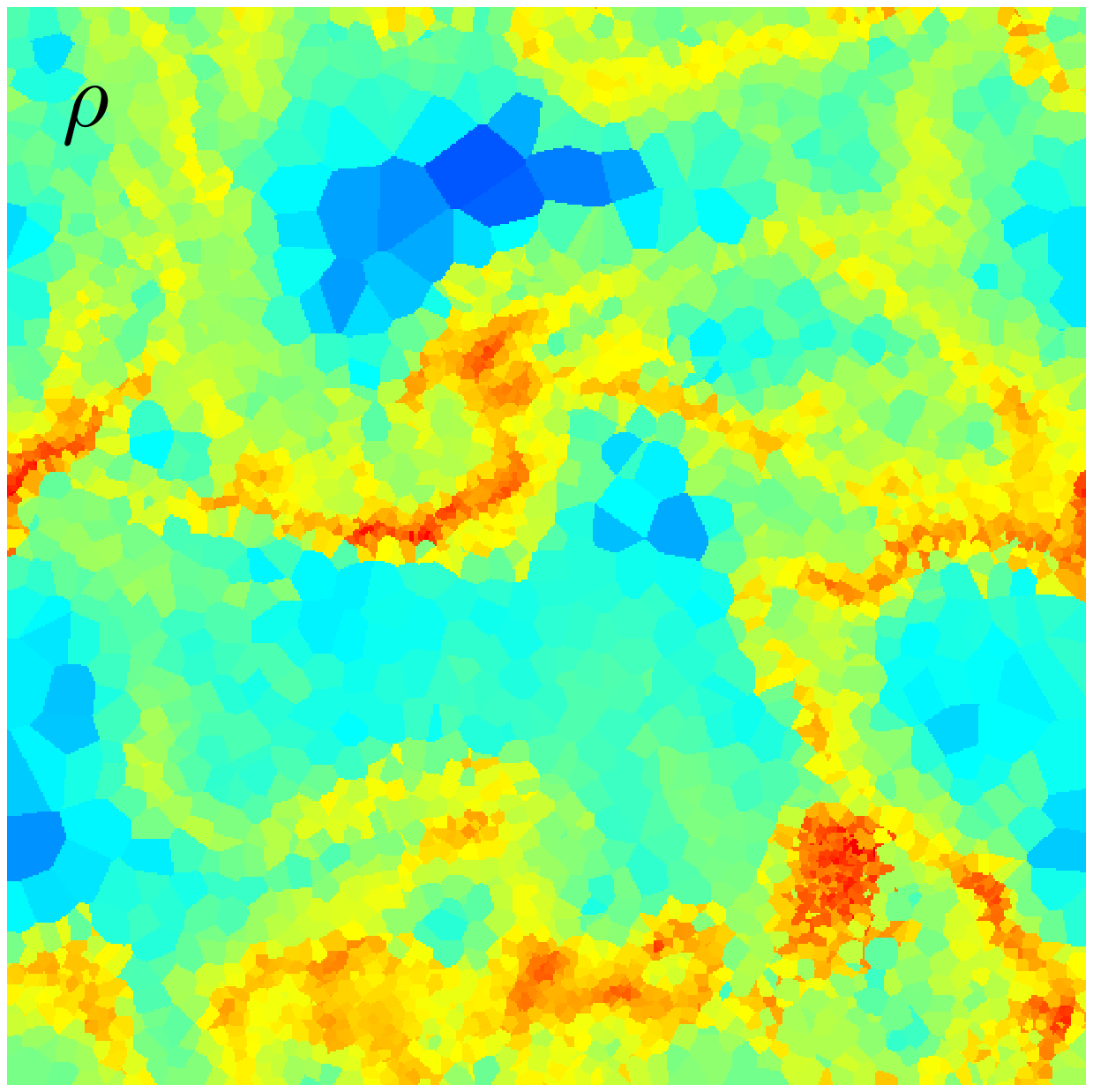}
&
\includegraphics[width=0.22\textwidth]{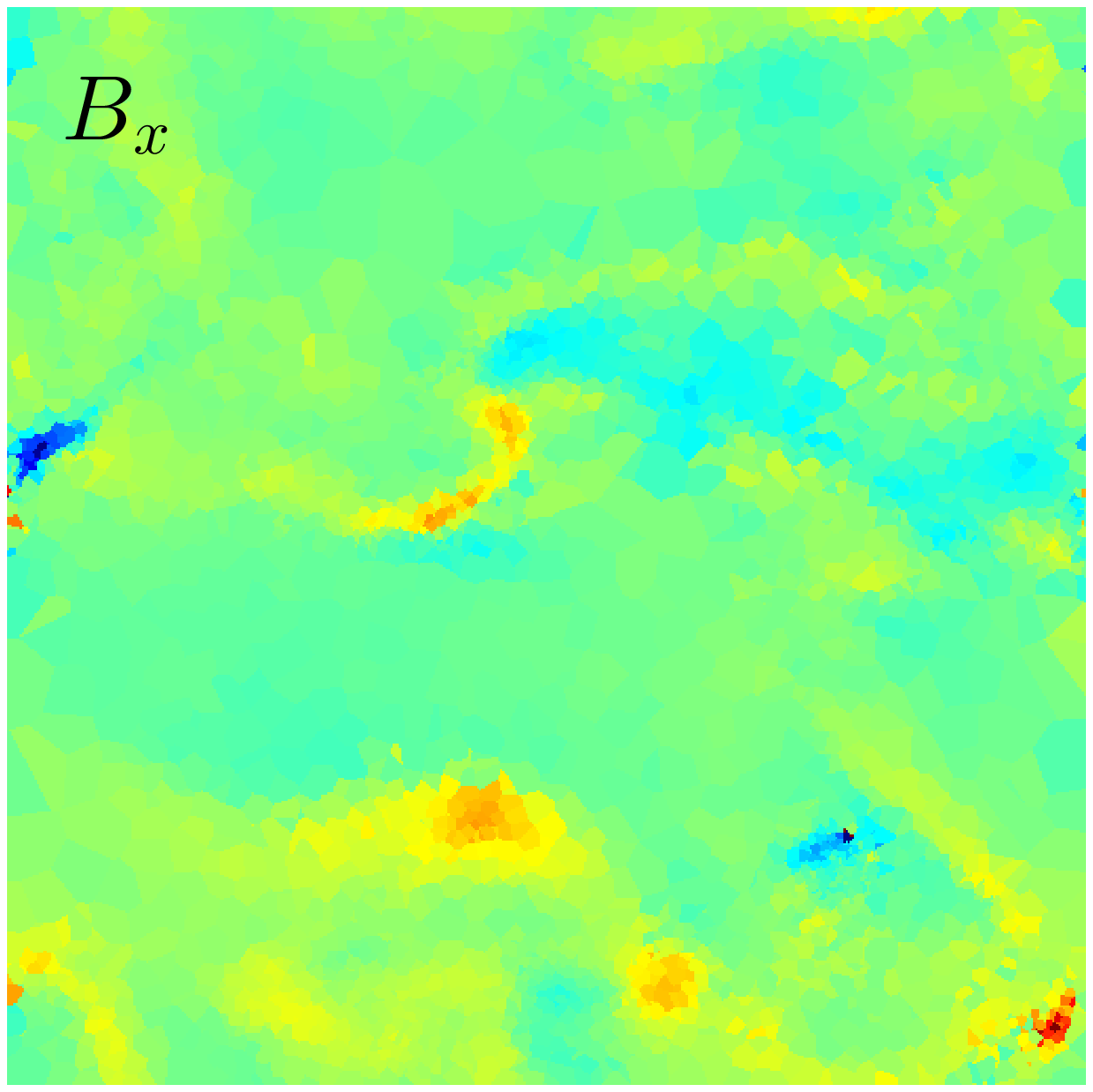}
&
\includegraphics[width=0.22\textwidth]{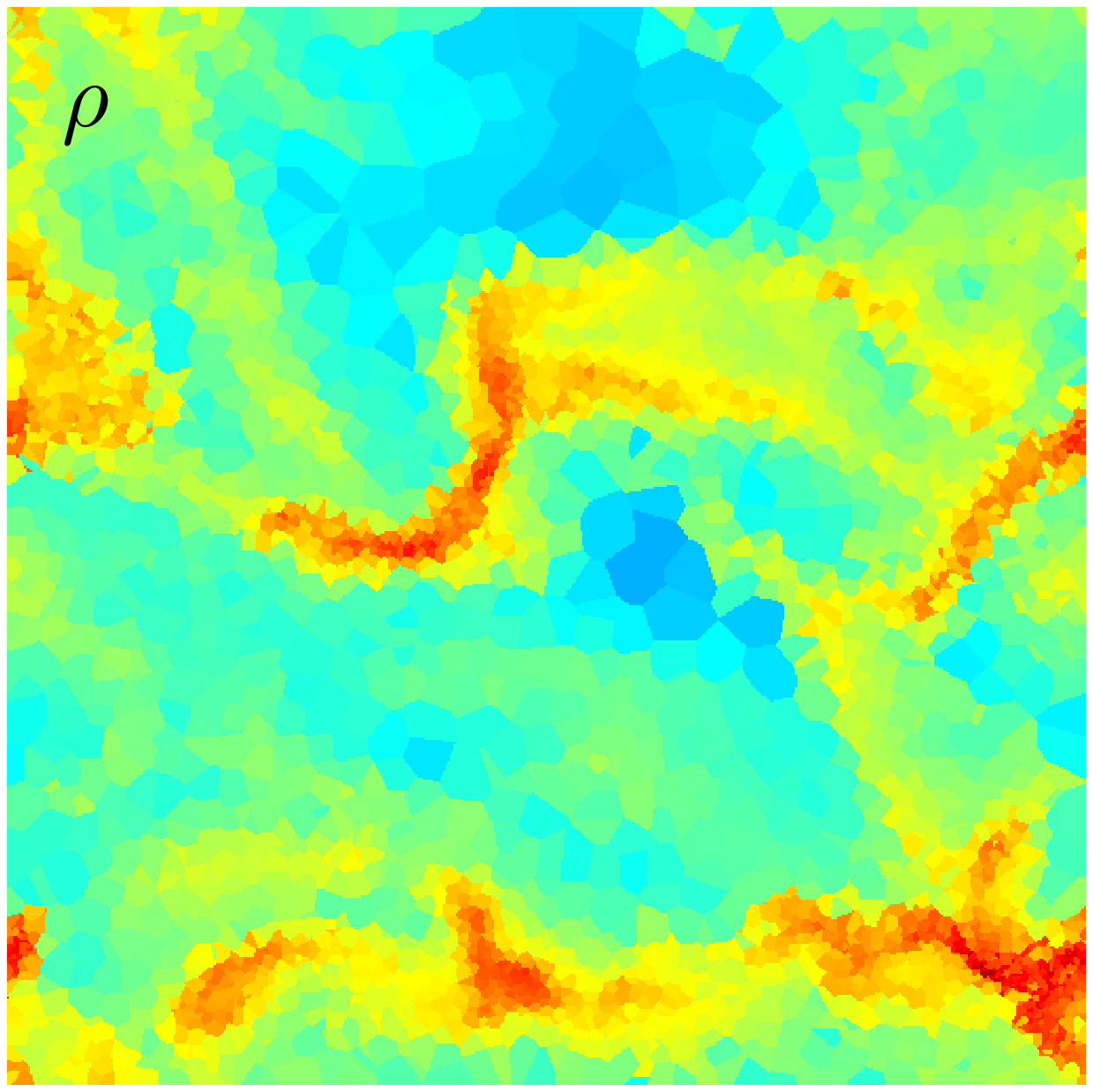}
&
\includegraphics[width=0.22\textwidth]{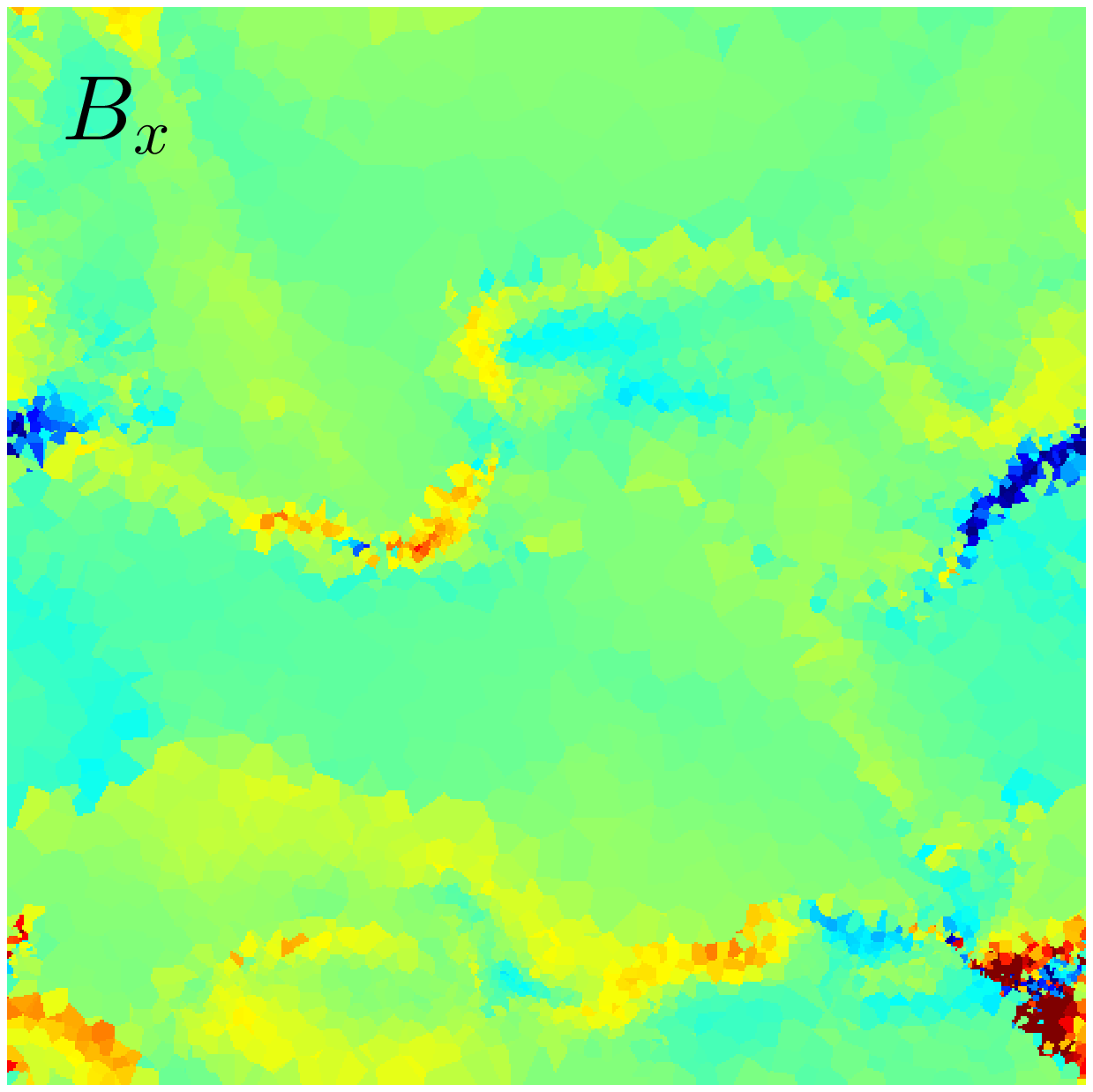}  \\
& & $\langle\mathbf{B}\rangle=[-0.01, 0.01, 0.97]$ & & 
$\langle\mathbf{B}\rangle=[0.06, 0.05, 1.21]$  \\
\begin{turn}{90} \qquad\qquad $t=10\tau_{\rm eddy}$ \end{turn}
&
\includegraphics[width=0.22\textwidth]{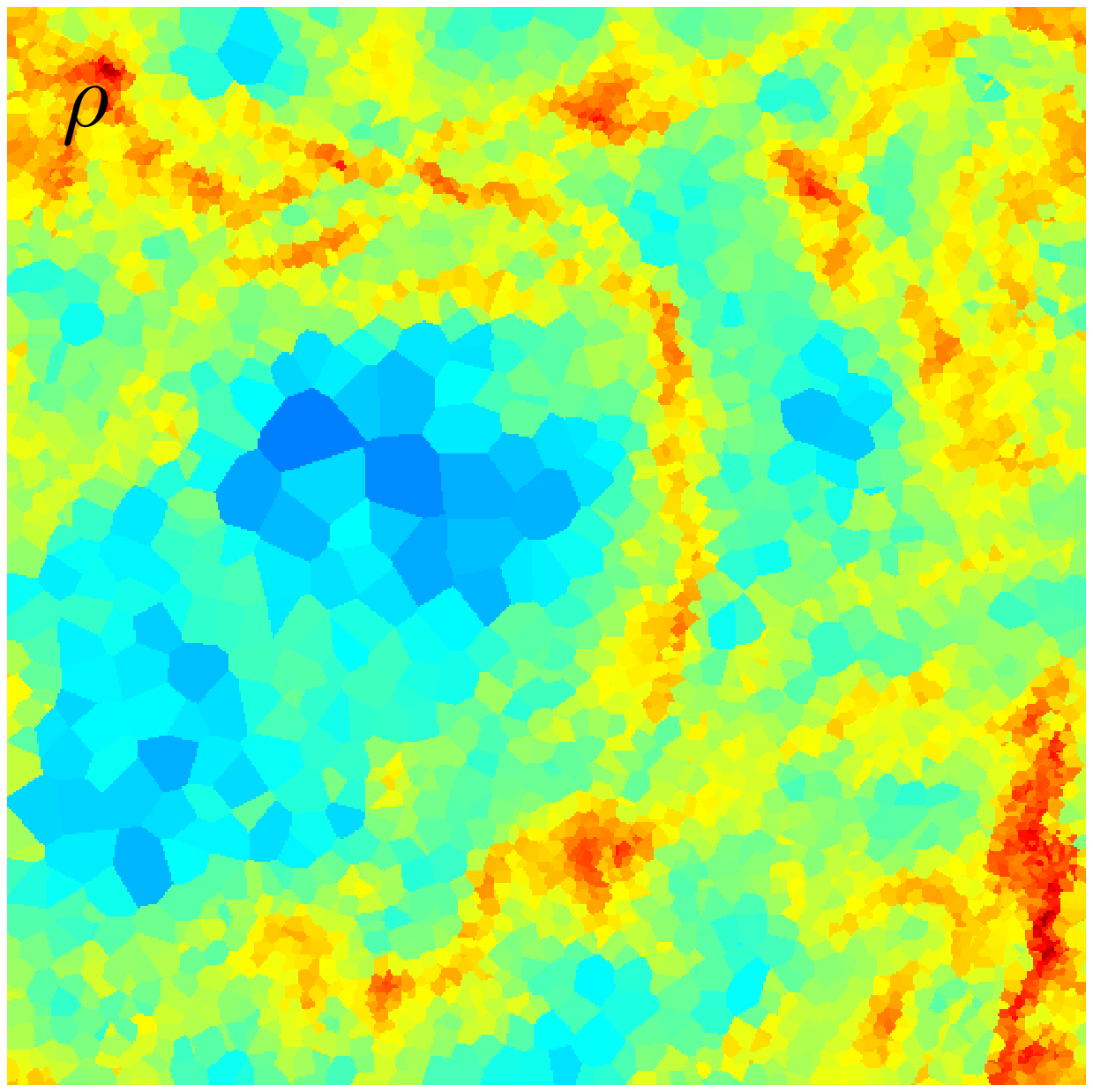}
&
\includegraphics[width=0.22\textwidth]{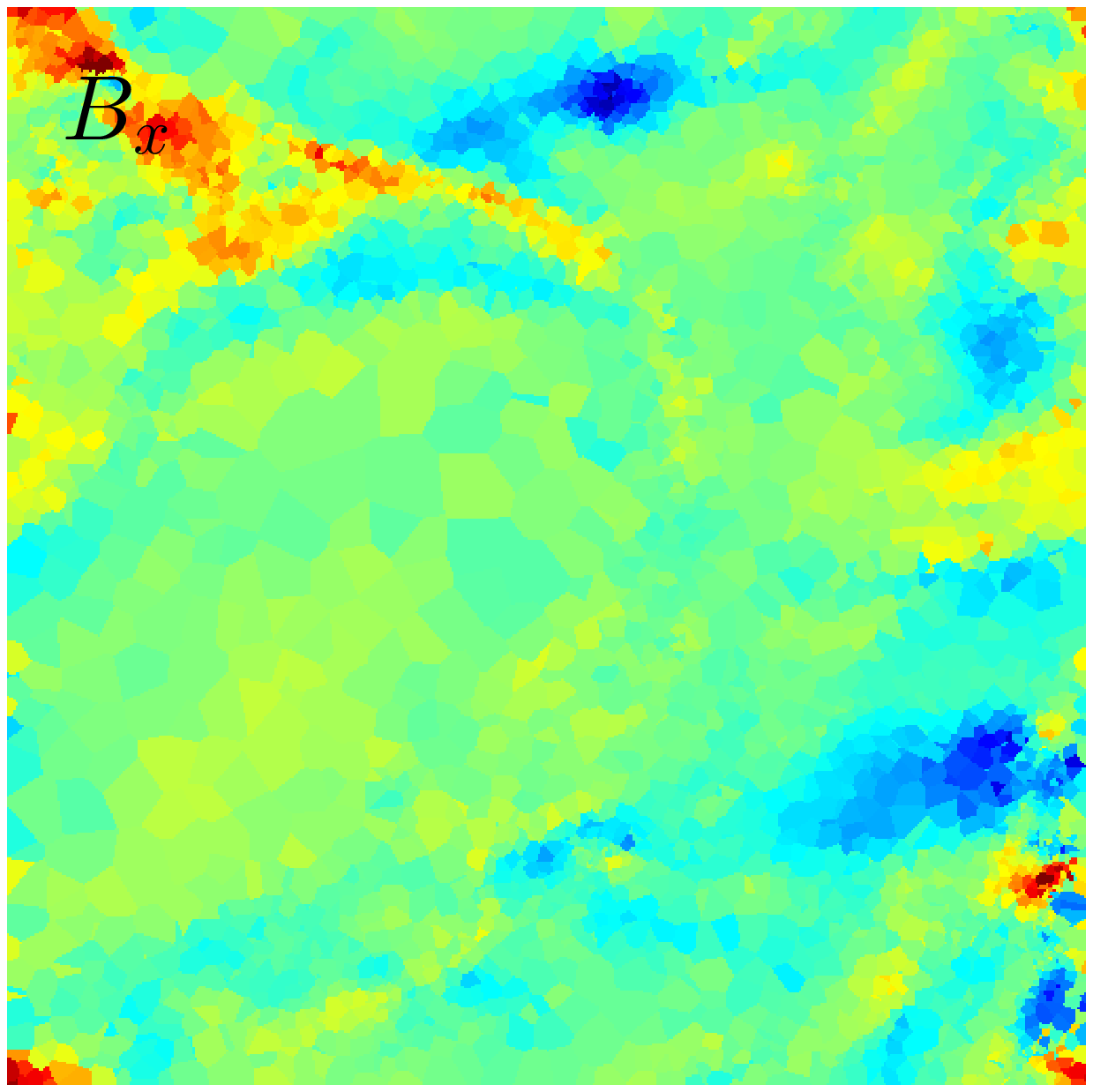}
&
\includegraphics[width=0.22\textwidth]{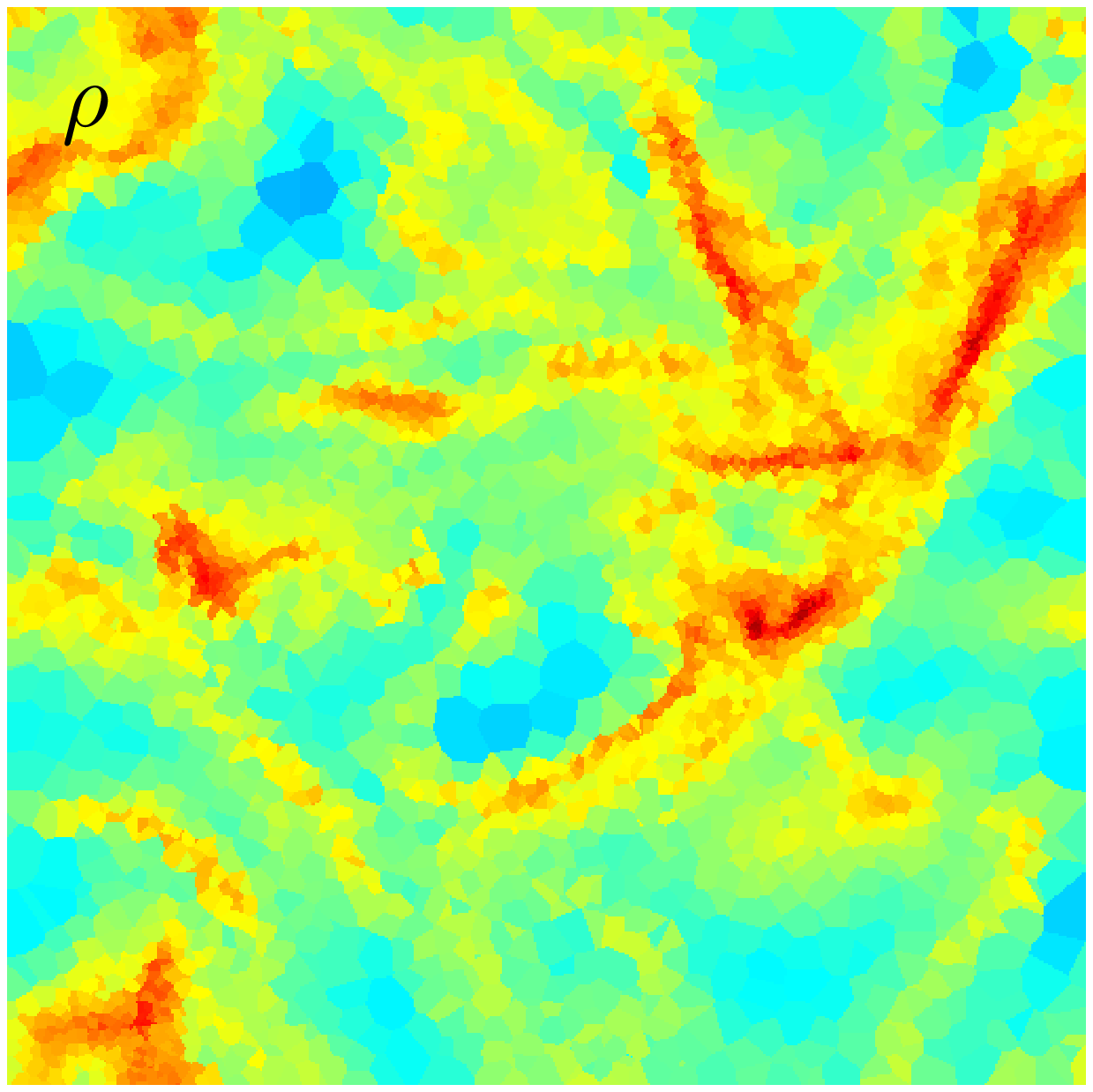}
&
\includegraphics[width=0.22\textwidth]{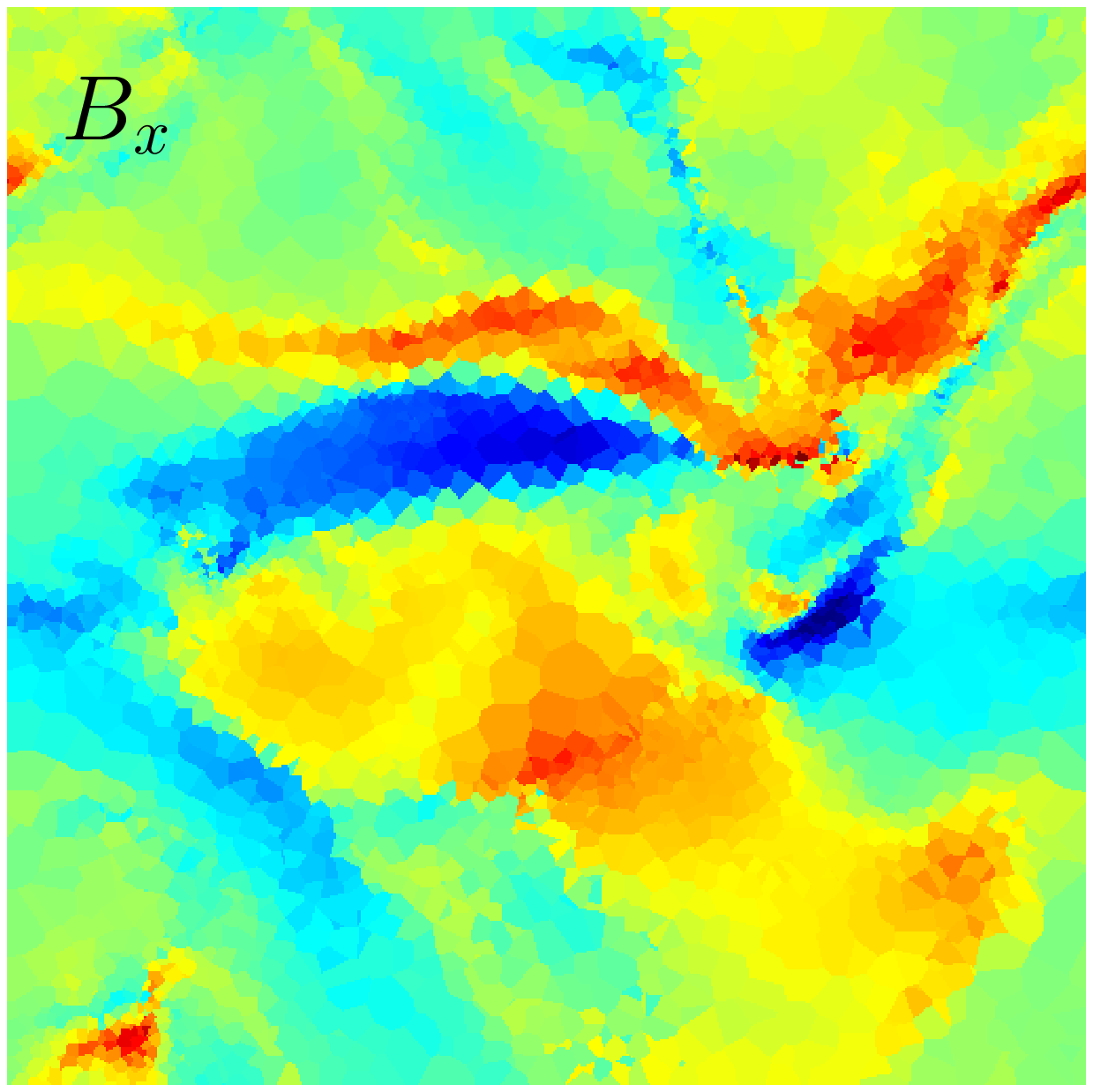}  \\
& & $\langle\mathbf{B}\rangle=[0.01, 0.00, 0.99]$ & & 
$\langle\mathbf{B}\rangle=[0.59, 1.46, 3.46]$  \\
\begin{turn}{90} \qquad\qquad $t=40\tau_{\rm eddy}$ \end{turn}
&
\includegraphics[width=0.22\textwidth]{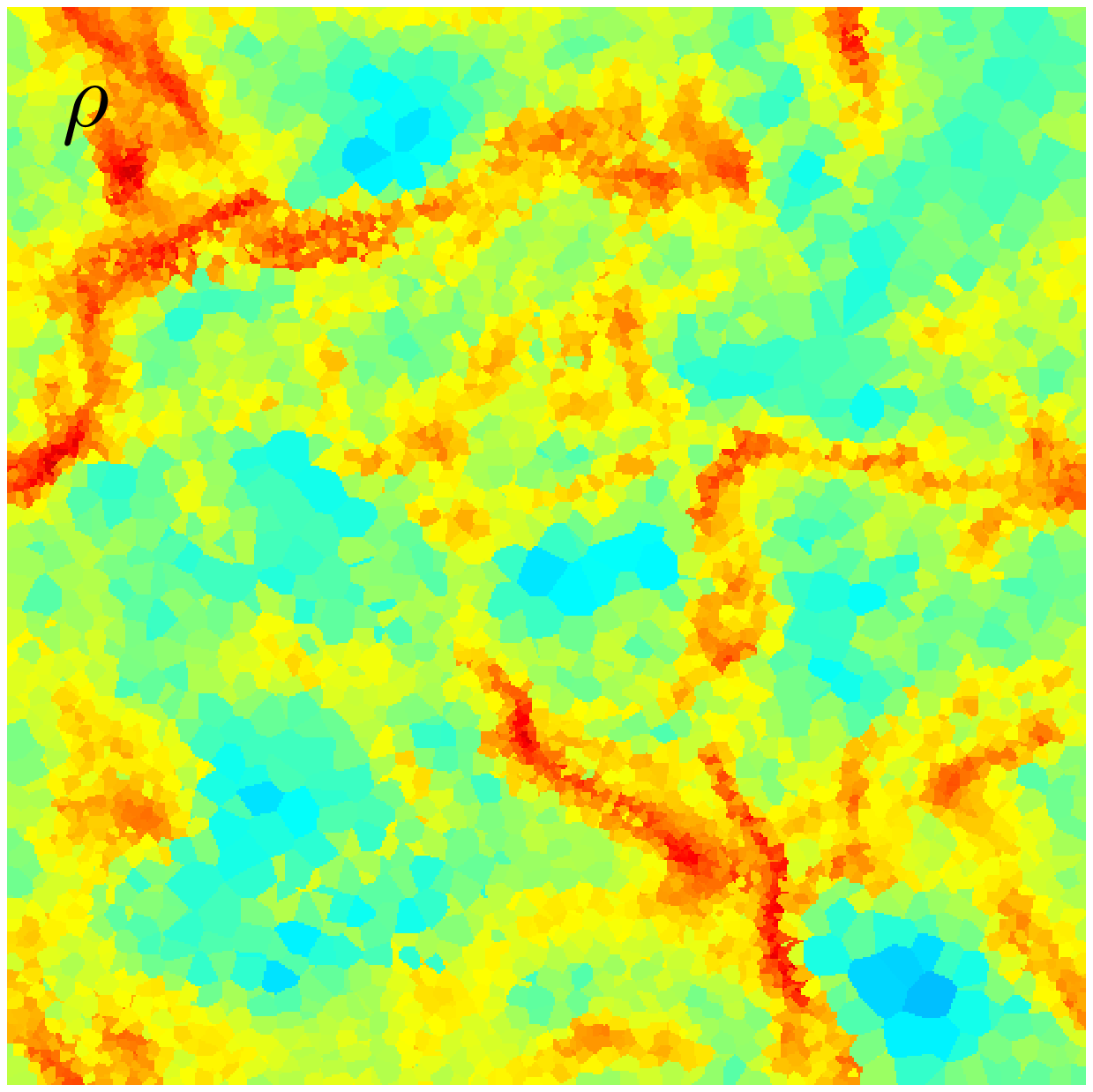}
&
\includegraphics[width=0.22\textwidth]{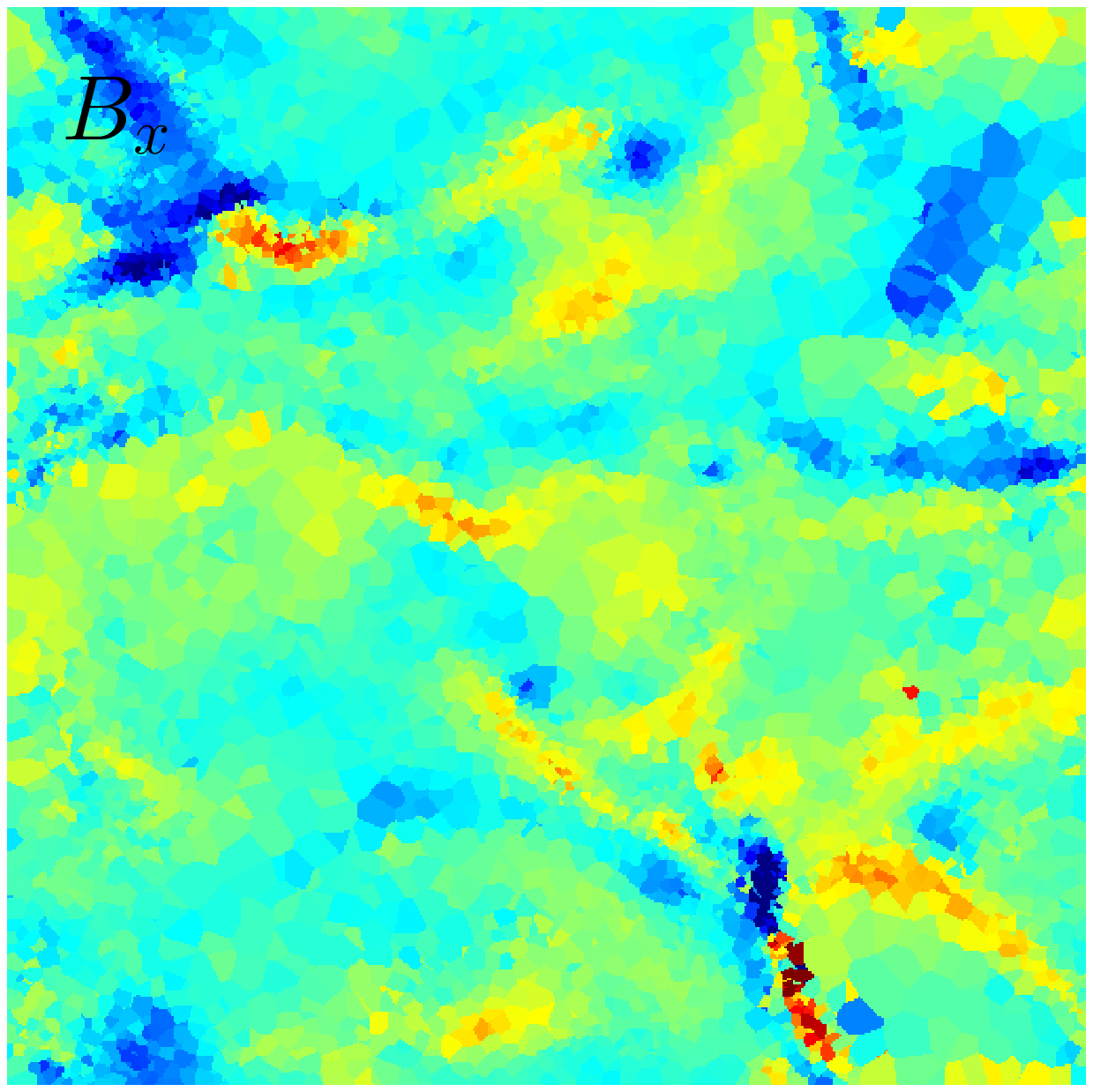}
&
\includegraphics[width=0.22\textwidth]{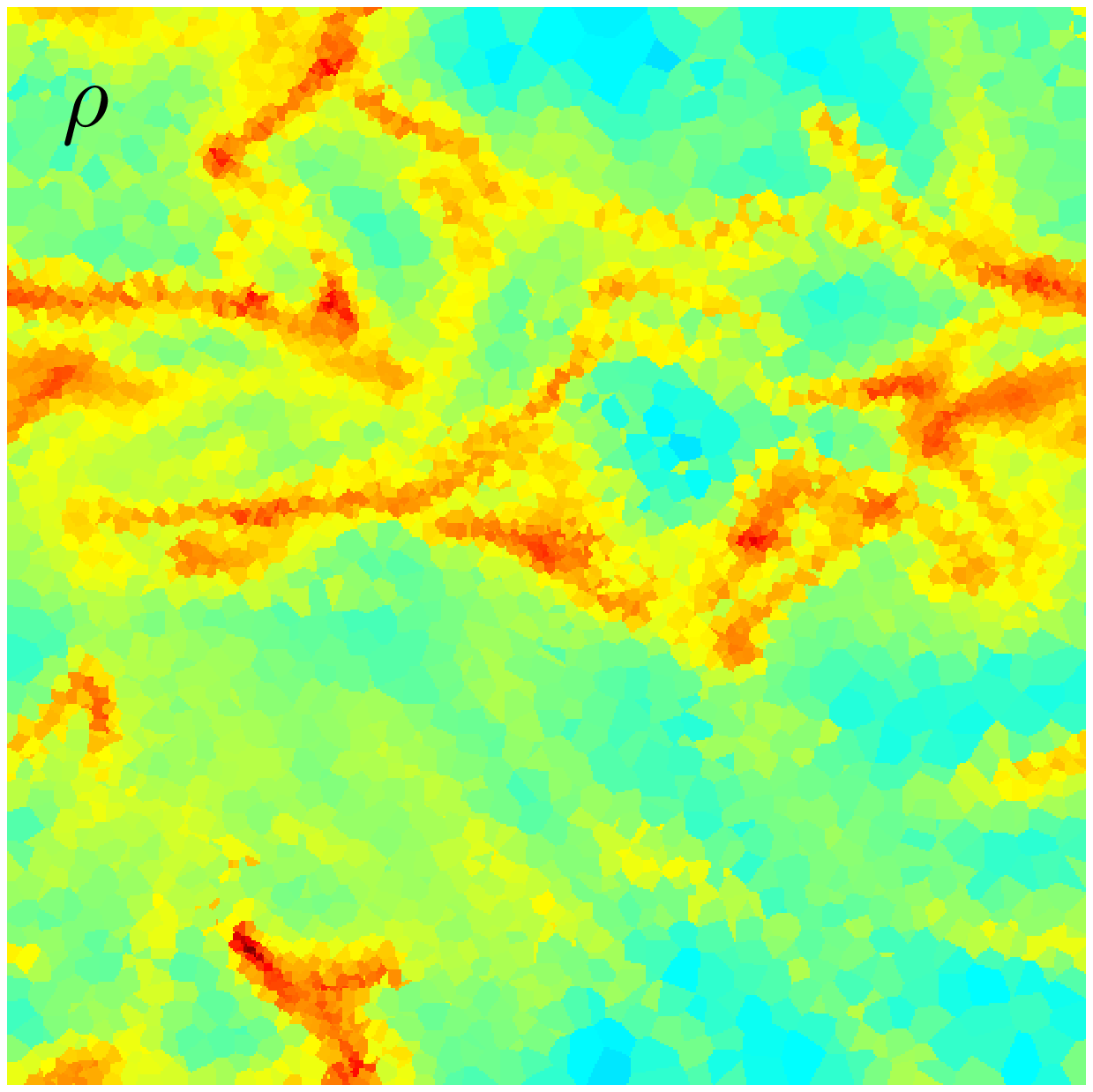}
&
\includegraphics[width=0.22\textwidth]{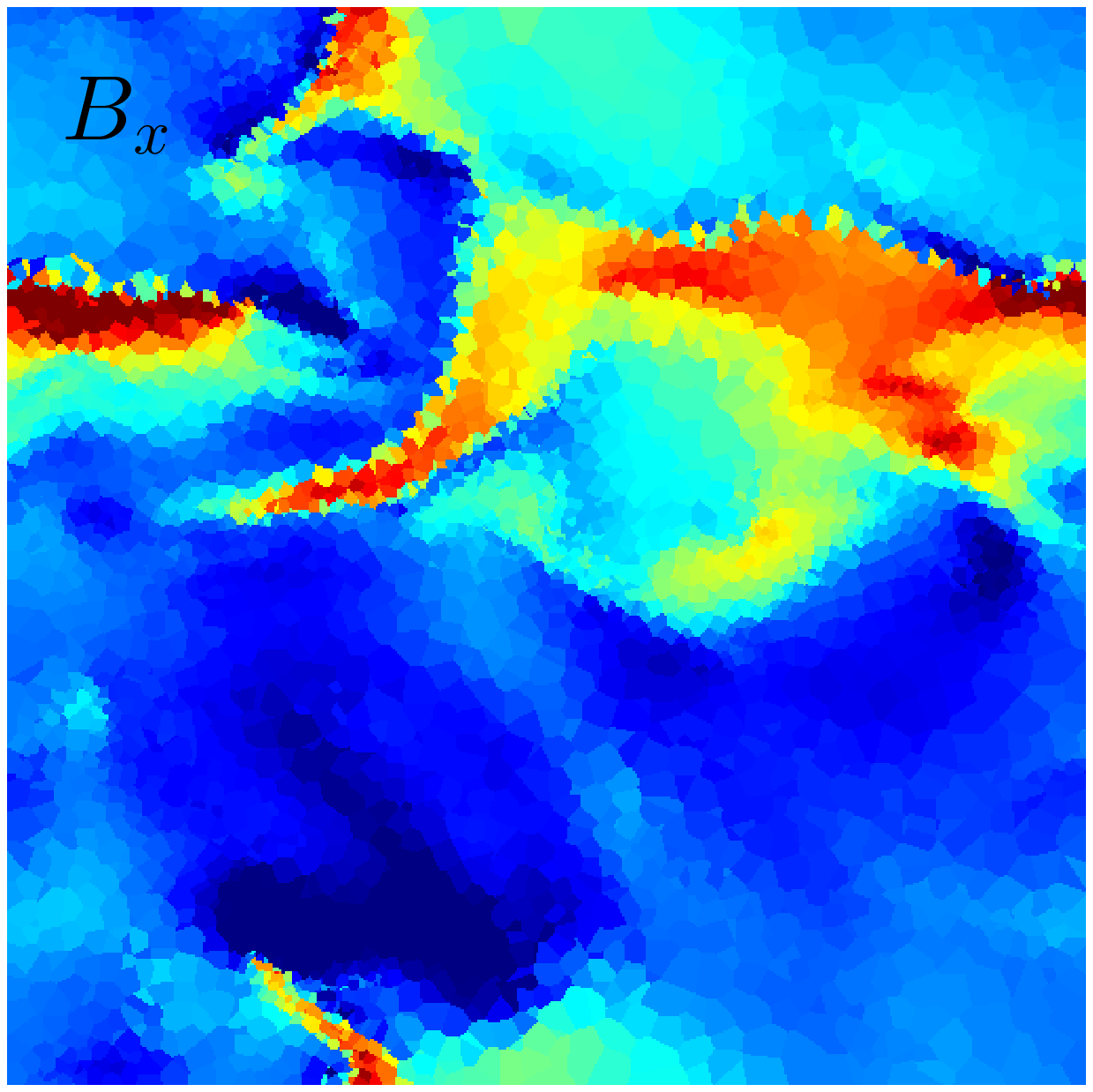}  \\
& & $\langle\mathbf{B}\rangle=[-0.01, -0.01, 0.99]$ & & 
$\langle\mathbf{B}\rangle=[-2.78, 7.63, 6.36]$  \\
&
\includegraphics[width=0.22\textwidth]{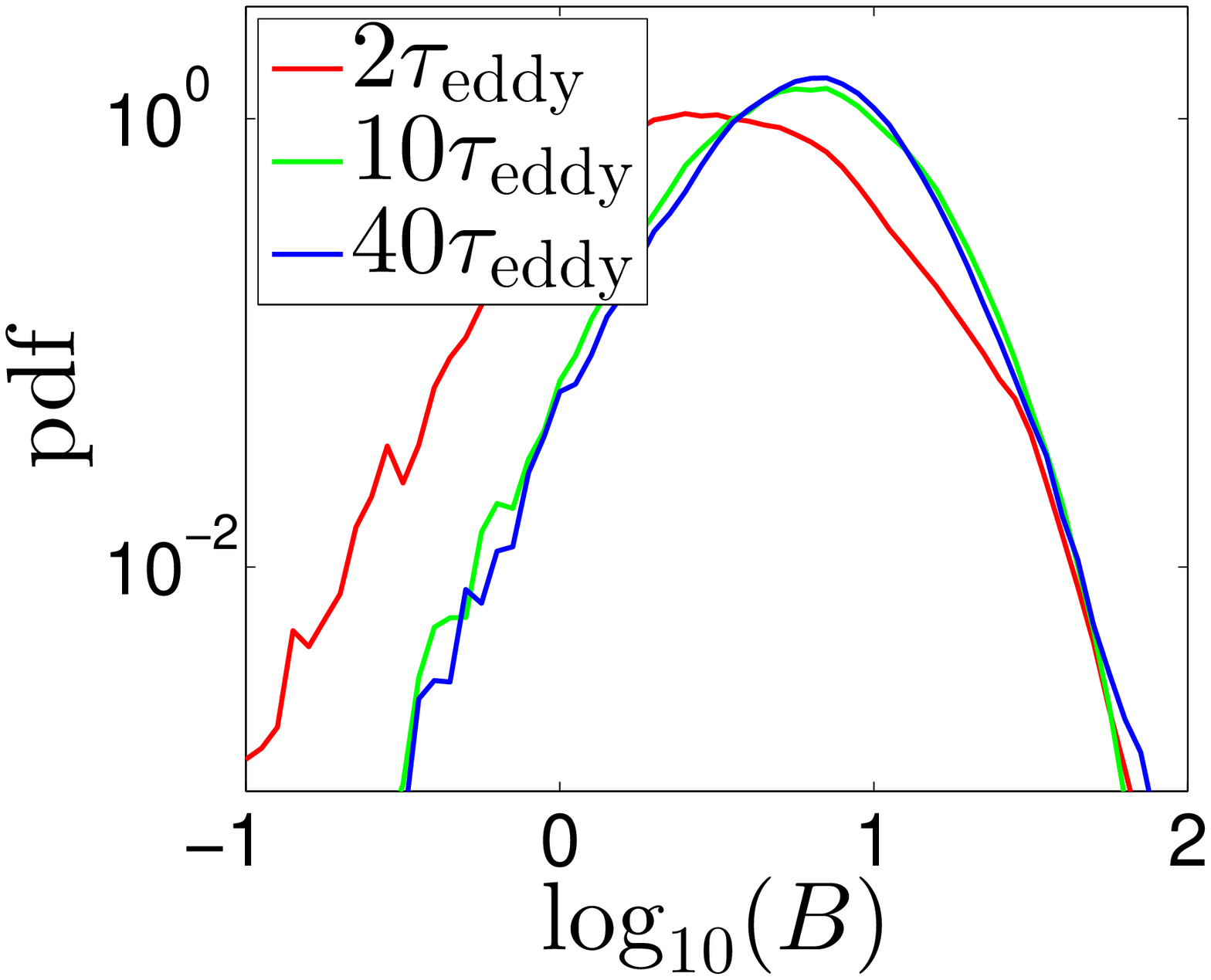}
&
\includegraphics[width=0.22\textwidth]{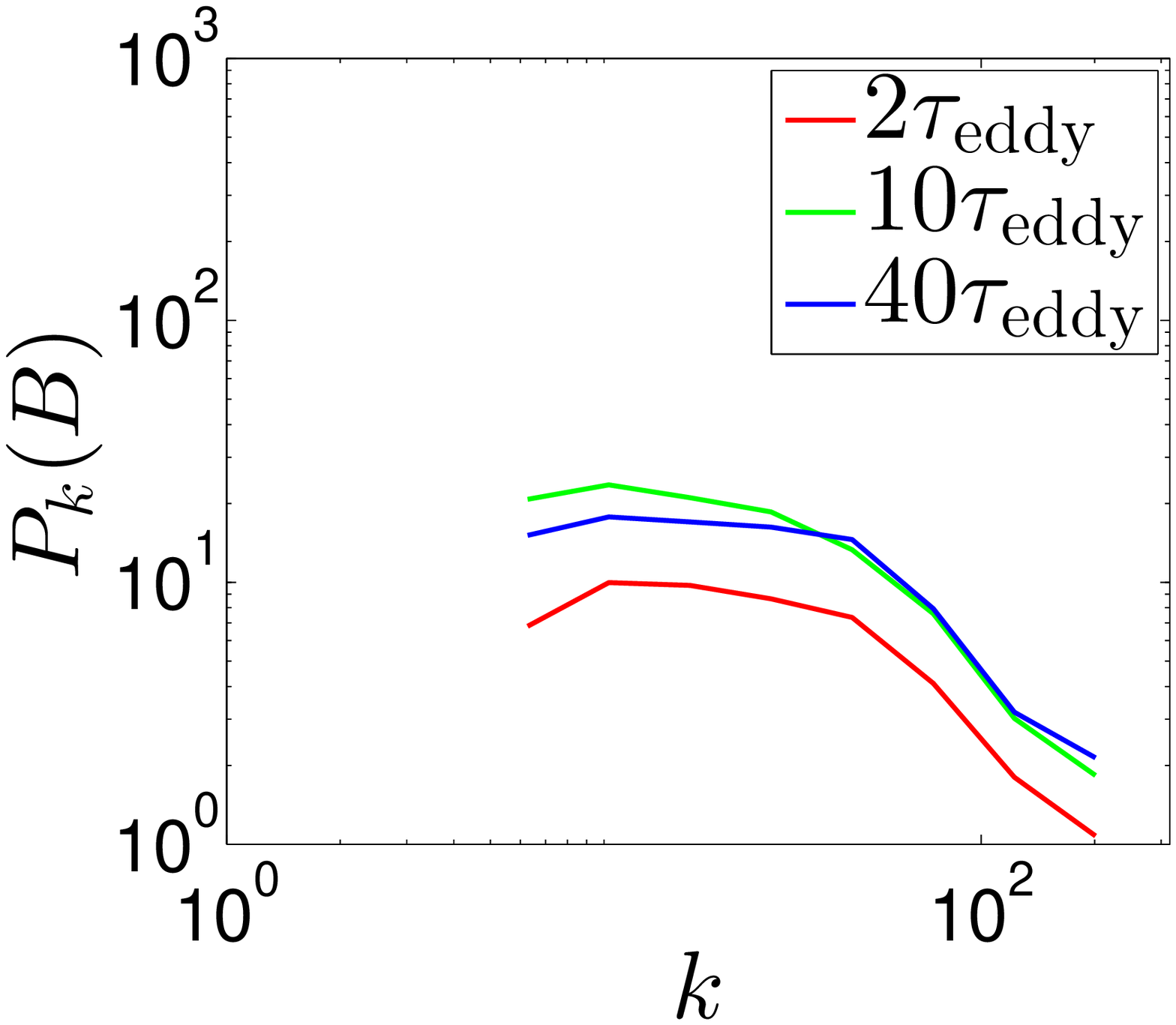}
&
\includegraphics[width=0.22\textwidth]{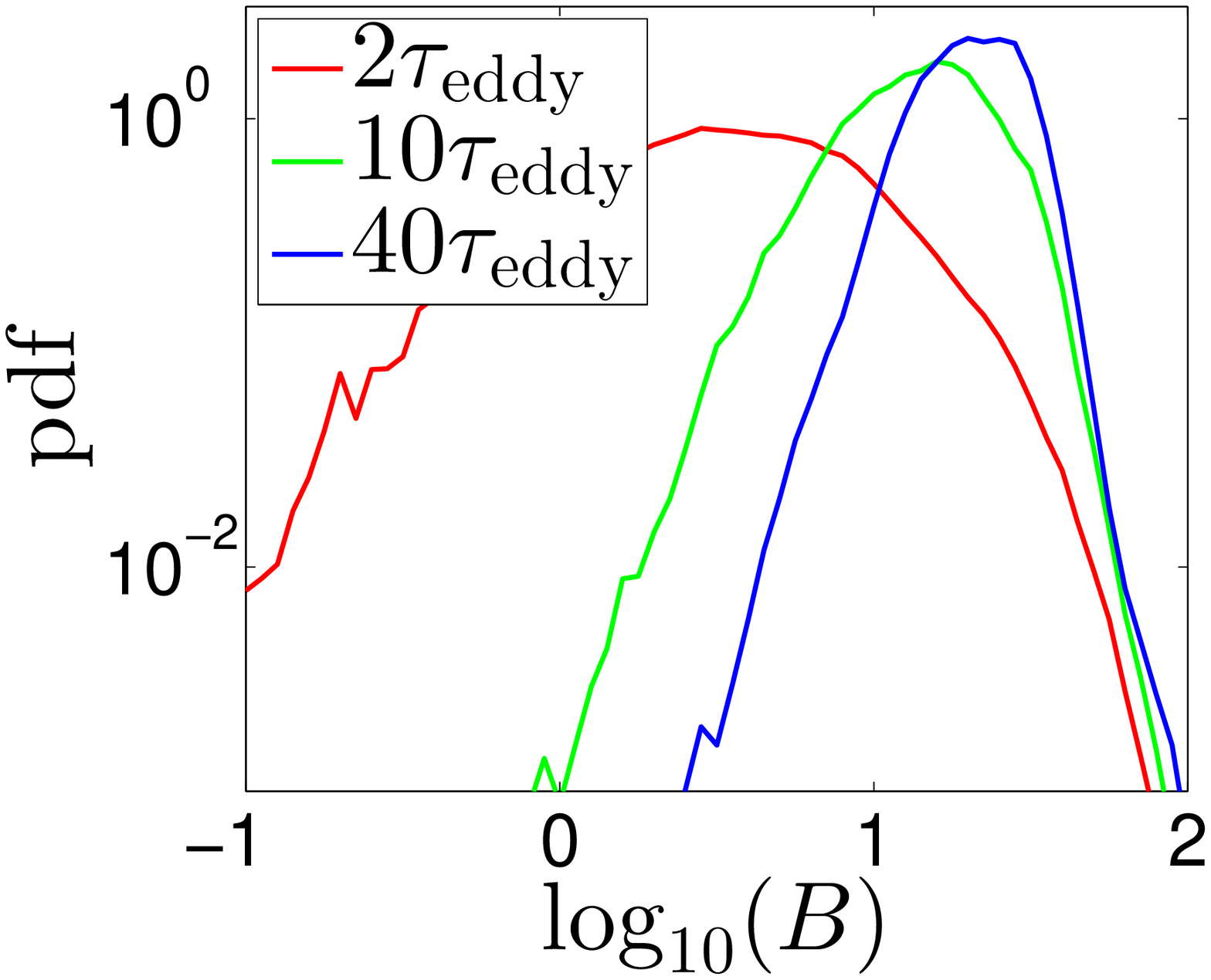}
&
\includegraphics[width=0.22\textwidth]{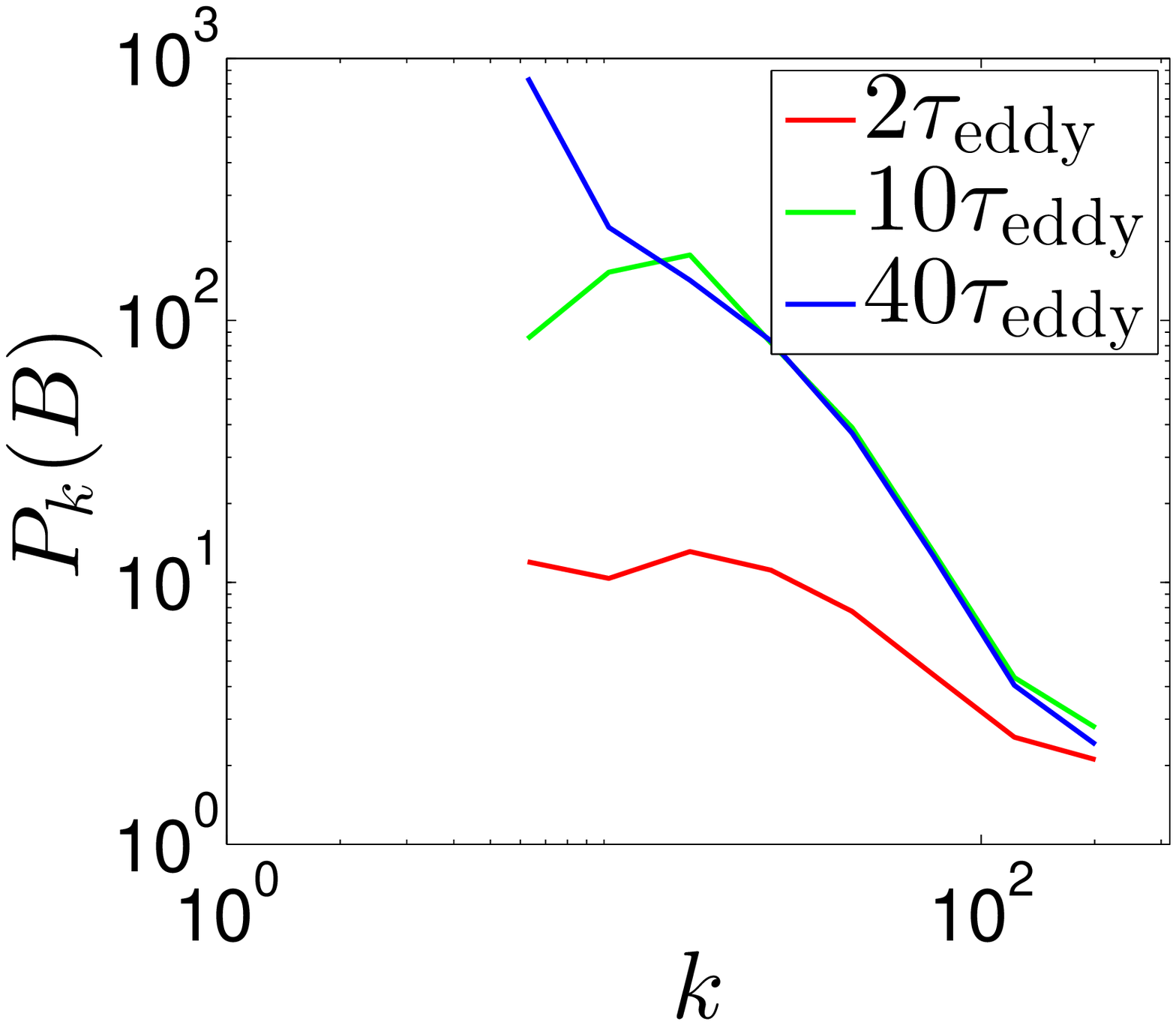}  \\
\end{tabular}
\caption{Comparison of the moving mesh CT and Powell schemes used to
simulate $\mathcal{M}_{\rm s}\sim 10$, $\mathcal{M}_{\rm A}\sim 3$ turbulence. Plotted are slices of the density field,
$x$ component of the magnetic field, and the
the PDF and power spectra of the magnetic field at $t=2,10,40$ eddy turnover times. In the plots of $B_x$ we also list the volume-averaged mean magnetic field in the domain, which is an ideal MHD invariant (its initial value was set to $\mathbf{B}=(0,0,1)$). At $2$ eddy turnover times, when the turbulence is transitioning into the non-linear saturated regime, the two schemes show similar features in the density and magnetic field. However, after more eddy turnovers, the Powell scheme artificially makes the mean magnetic field grow significantly due to its non-conservative formulation and shifts the entire magnetic field PDF of the magnetic field to the right, as well as transfers most of the magnetic energy to the largest scale ($k=2\cdot\pi$). The CT scheme offers good control over the mean magnetic field and shows a stable PDF and power spectrum of the magnetic field.
}
\label{fig:turb}
\end{figure*}

\begin{figure*}
\centering
\begin{tabular}{ccc}
 \large{moving CT}  & \large{moving Powell} & \\
\includegraphics[width=0.29\textwidth]{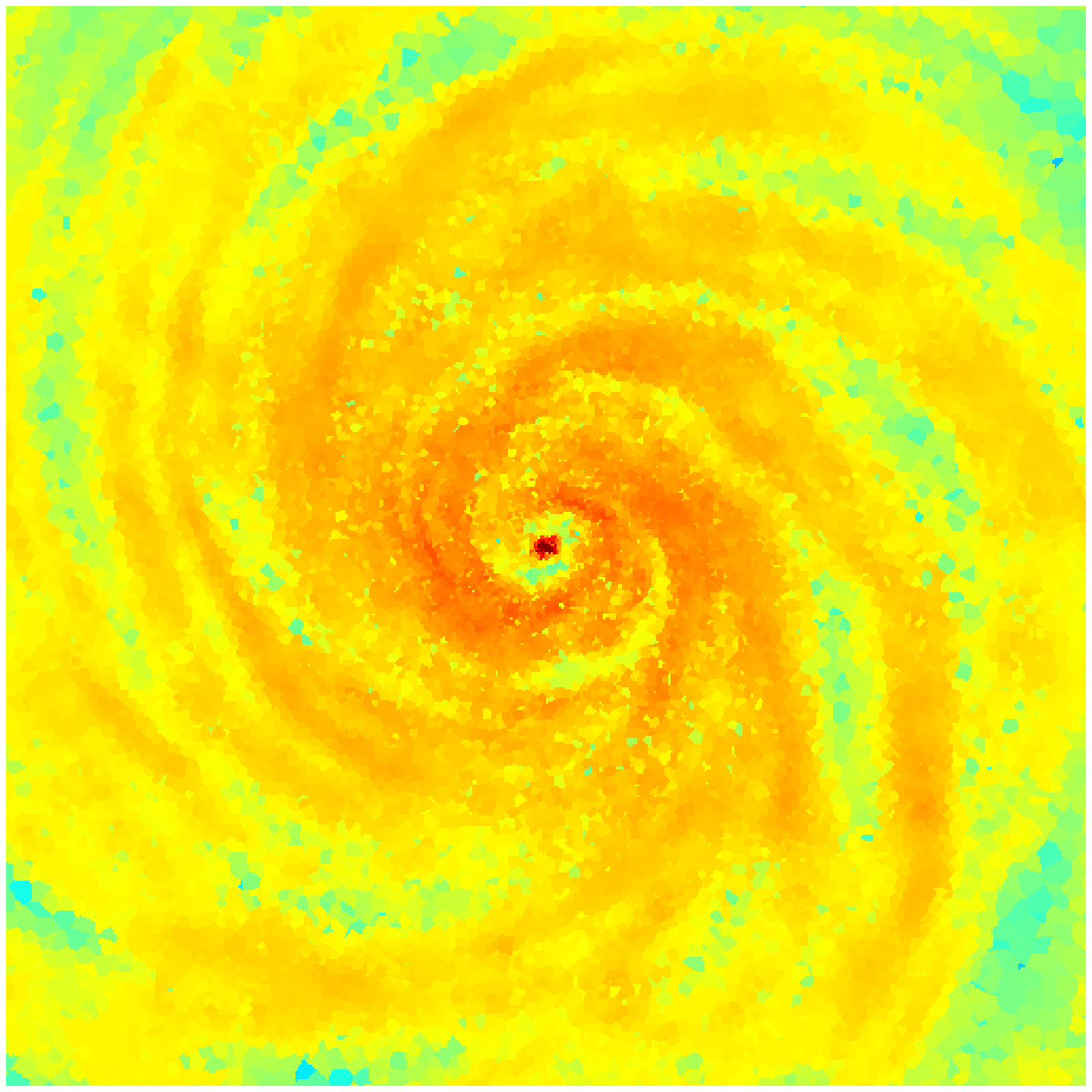} &
\includegraphics[width=0.29\textwidth]{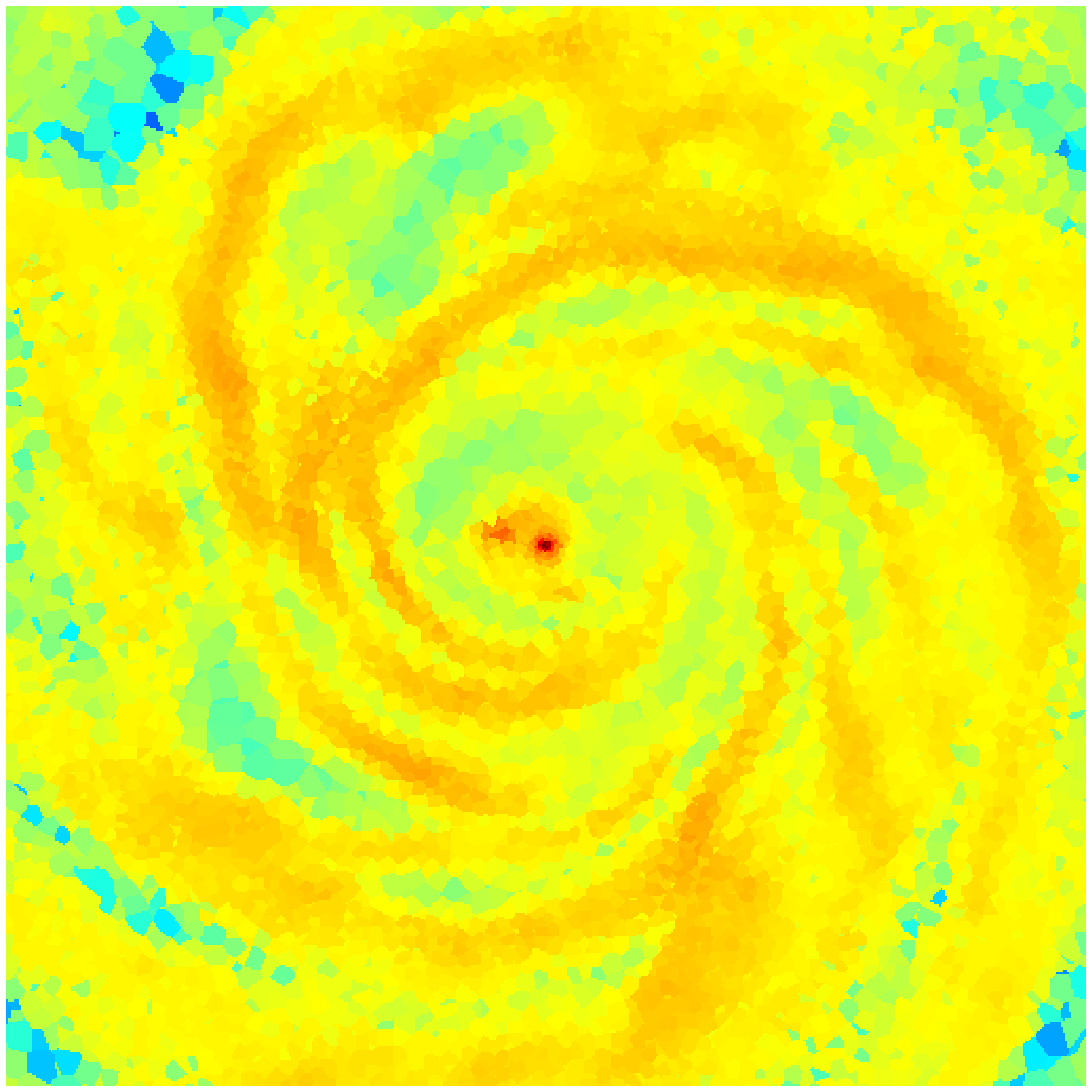} &
\includegraphics[width=0.056\textwidth]{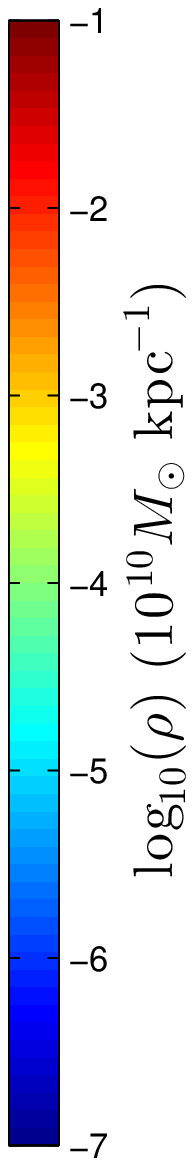} \\
\includegraphics[width=0.29\textwidth]{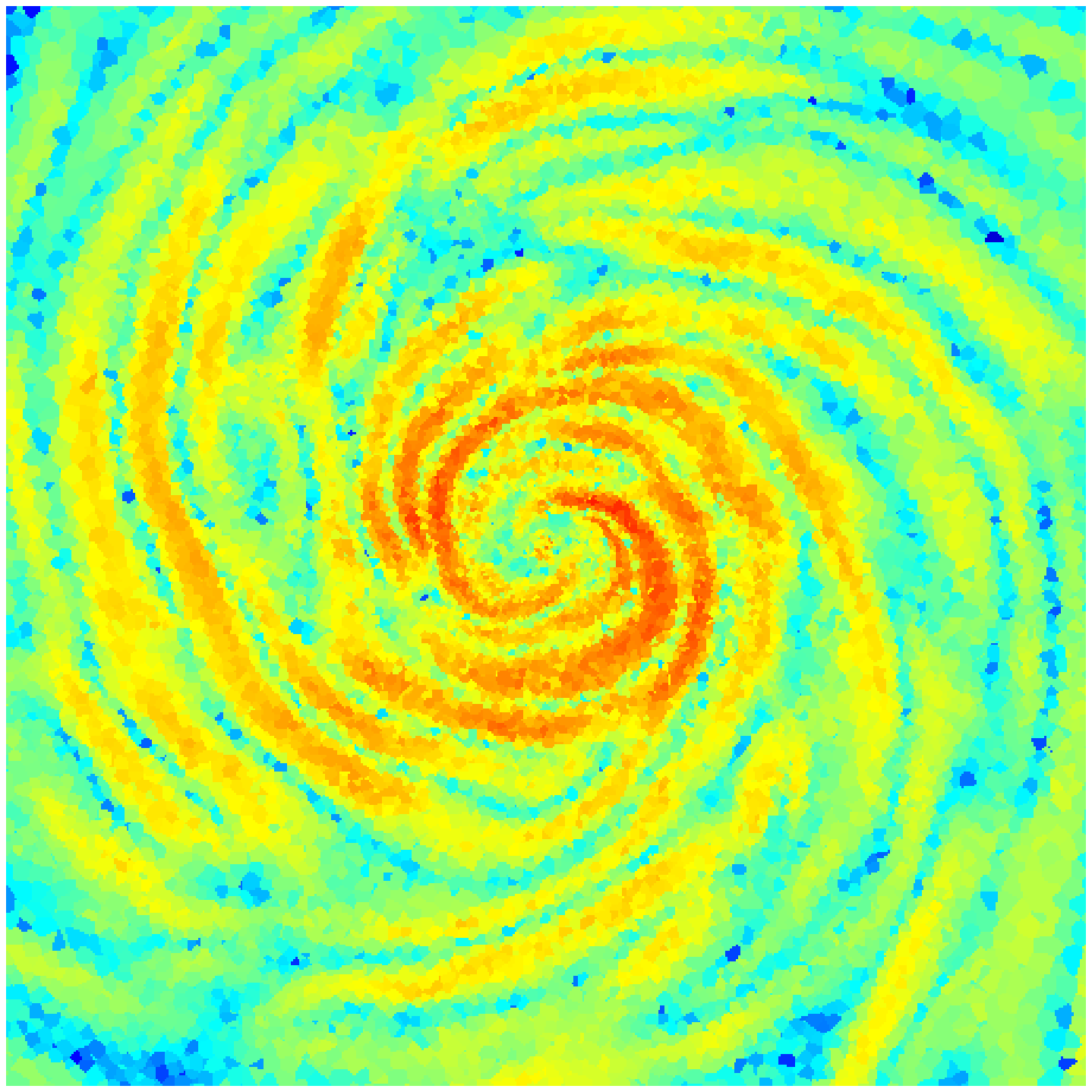} &
\includegraphics[width=0.29\textwidth]{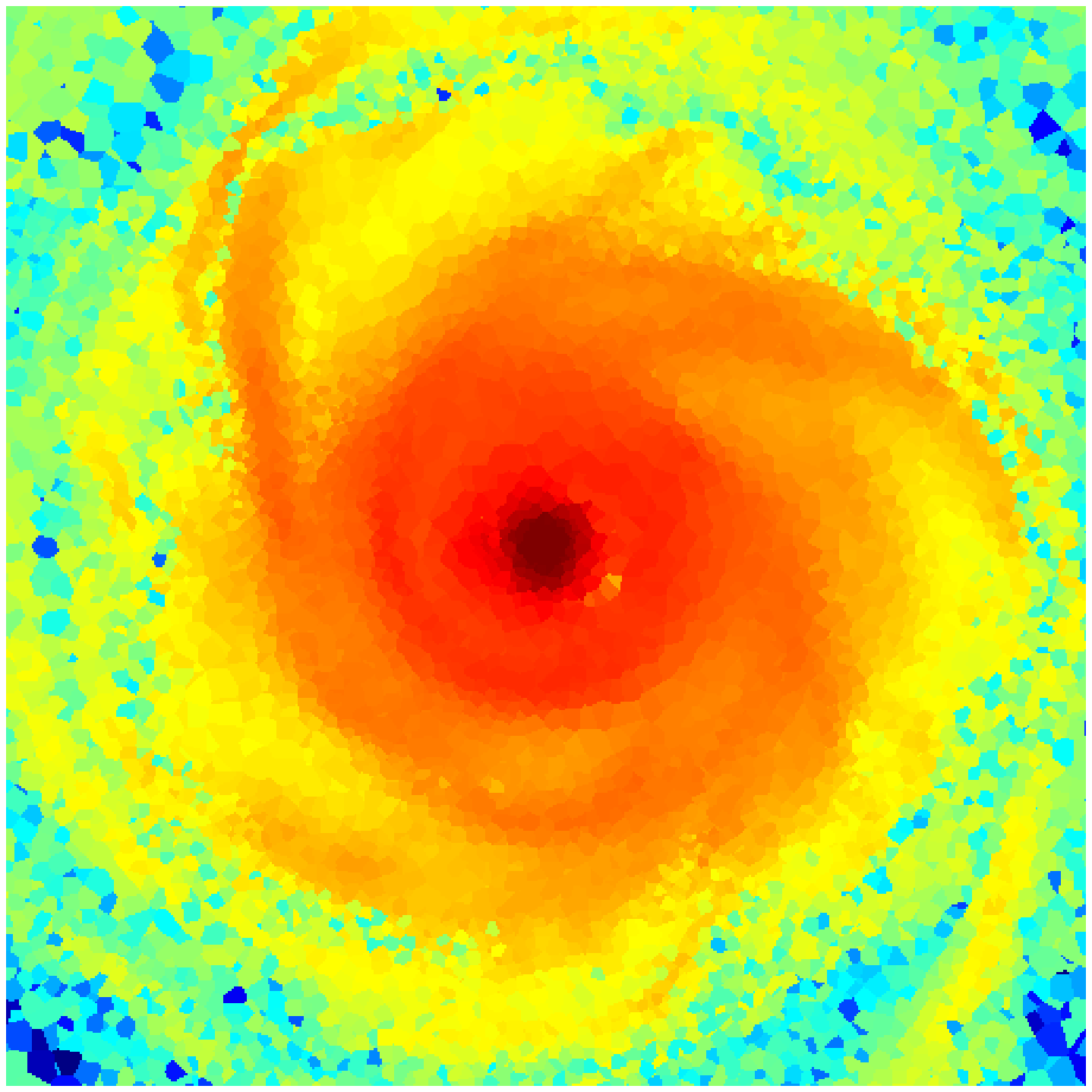} &
\includegraphics[width=0.056\textwidth]{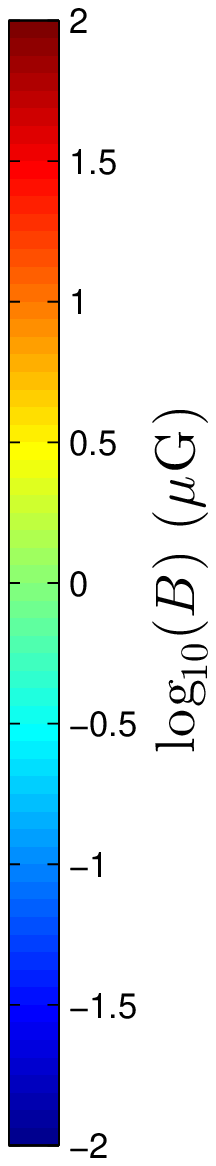} \\
\includegraphics[width=0.29\textwidth]{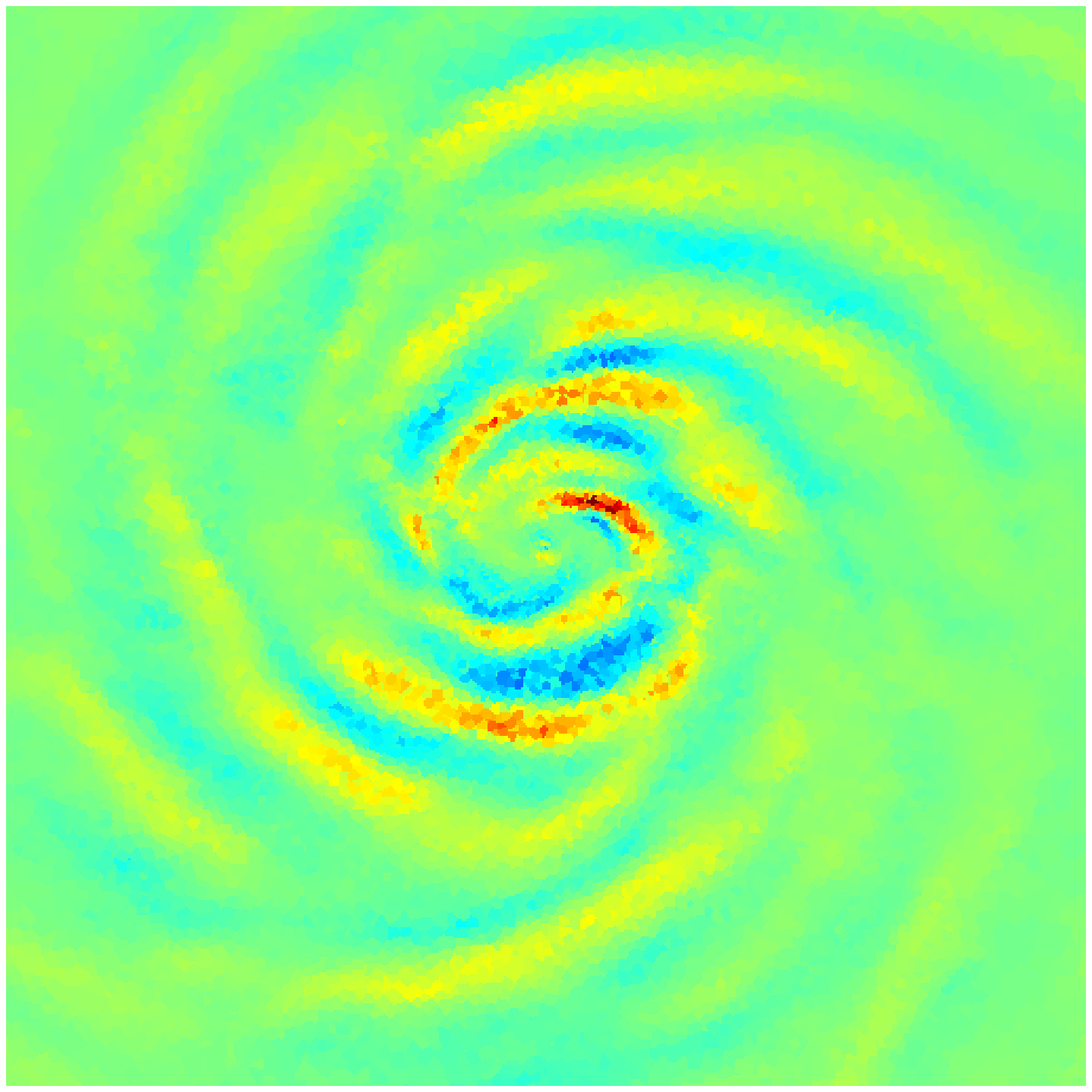} &
\includegraphics[width=0.29\textwidth]{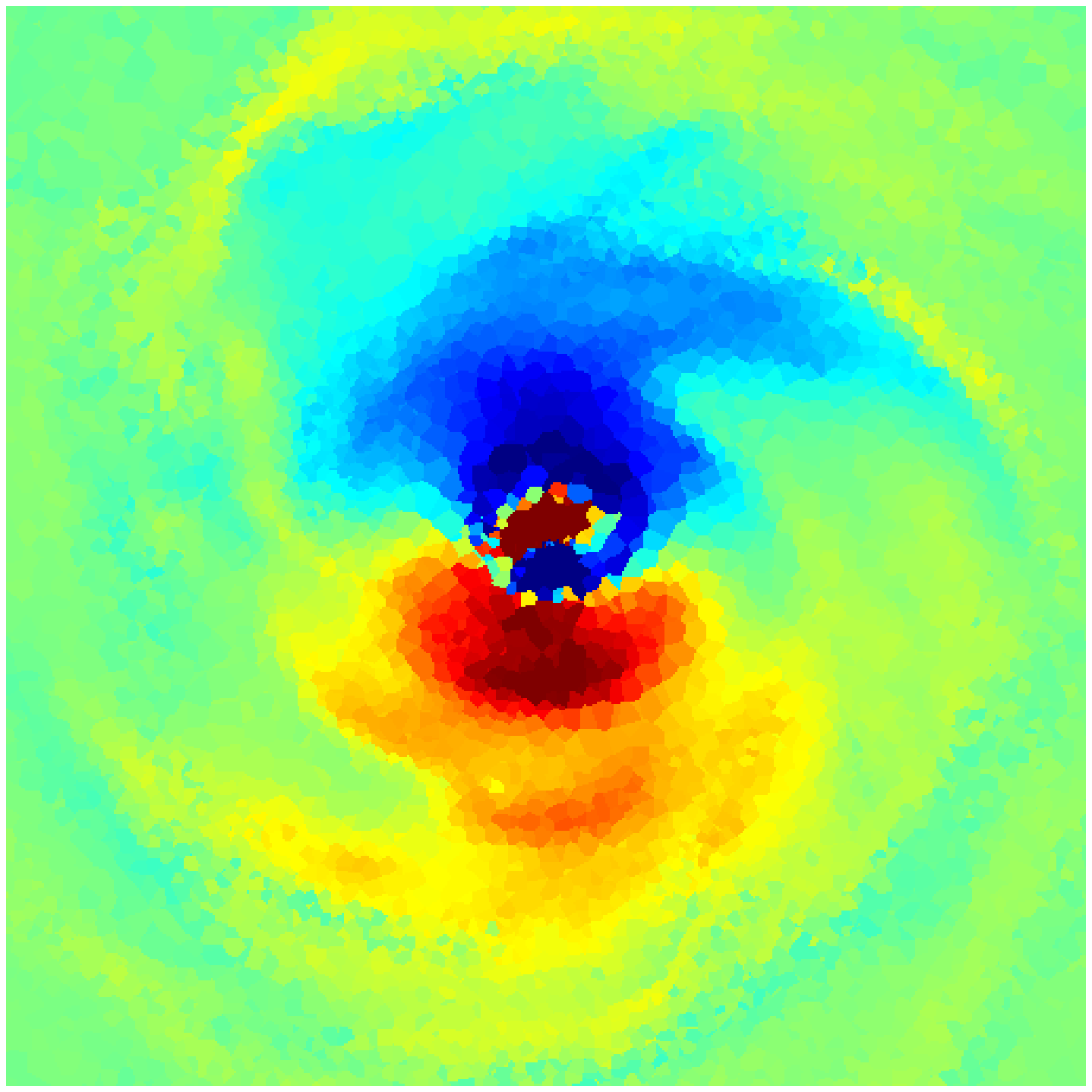} &
\includegraphics[width=0.056\textwidth]{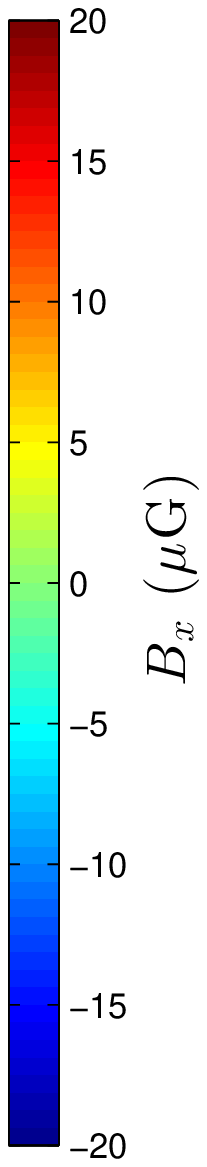} 
\end{tabular}
\caption{Comparison of magnetic field saturation in the formation
of a disc simulated with the CT and Powell schemes.  Shown are the density, magnetic field strength, and magnetic field $x$
component, of the disc at $t=2.5$~Gyr. Each
panel displays a physical region of side length $40~{\rm kpc}$. With the
Powell scheme, unlike the CT method, the magnetic field has grown
dynamically dominant and disrupts the disc.}
\label{fig:disc}
\end{figure*}

The CT mapping is as follows. Define the modified Delaunay
tetrahedralization, by which we mean the tetrahedralization obtained
from the Delaunay connections, but with the nodes, originally located
at the mesh generating points, shifted slightly to the centres-of-mass
of the cells.  We will refer to the tetrahedra in this
tetrahedralization as Delaunay tetrahedra but note that their nodes
are slightly offset from the standard definition. Using the
centres-of-mass instead of the mesh generating points is formally more
accurate for our scheme, but the method may be implemented using the
mesh generating points instead, with little effect on the results on a
well-regularized mesh.

[1] First, the components $\mathbf{A}_i$ are projected onto each of
the connections in a modified Delaunay tetrahedralization of the
mesh. That is, we obtain
\begin{equation}
\mathcal{A}_{ij} = \mathbf{A}_i\cdot d\mathbf{r}_{ij}
\end{equation}
where $\mathbf{r}_{ij}$ is the connection pointing from the
center-of-mass of cell $i$ to the center-of-mass of cell $j$.

[2] Next, the outward normal component of the magnetic flux
$\Phi_{ijk}$ is computed on each of the Delaunay tetrahedral faces
(with vertices labelled by mesh generating point $i$, $j$, and
$k$). We note that in an unstaggered CT approach, such as that of
\cite{2014MNRAS.442...43M}, the quantities $\Phi_{ijk}$ would be
evolved directly. However, evolving the vector potential instead and
mapping down to face fluxes is more efficient and easier to
implement. The $\Phi_{ijk}$ pointing along the direction given by the
right-hand rule with nodes $i$, $j$, and $k$ counter-clockwise, is:
\begin{equation}
\Phi_{ijk} = \frac{1}{2}\left( \mathcal{A}_{ij}-\mathcal{A}_{ji} + \mathcal{A}_{jk}-\mathcal{A}_{kj} + \mathcal{A}_{ki} - \mathcal{A}_{ik} \right)
\end{equation}
as it comes from the equation for the magnetic flux through a surface 
\begin{equation}
\Phi_S = \oint_{\partial S} \mathbf{A}\cdot d\mathbf{\ell}
\end{equation}
where the line integral is computed with a counter-clockwise
orientation around the boundary of the surface $S$.

[3] Next, the full magnetic field $\mathbf{B}_{ijkl}$ in each Delaunay
tetrahedron is reconstructed from the magnetic fluxes $\Phi_{ijk}$,
$\Phi_{ijl}$, $\Phi_{jkl}$, $\Phi_{kil}$. Note that the divergence
free condition implies $\Phi_{ijk}+\Phi_{ijl}+\Phi_{jkl}+\Phi_{kil} = 0$
(it is exactly this value that is preserved to machine precision by
our CT approach). So there are $3$ degrees of freedom represented in
both $\mathbf{B}_{ijkl}$ and the face magnetic fluxes. Hence, there is
a unique magnetic field vector that projects onto the faces to give
the four desired (divergence-free) fluxes. We find the value of the
magnetic field by inverting the linear system
\begin{equation}
\begin{pmatrix}
\mathscr{A}_{x,ijk}  \,  \mathscr{A}_{y,ijk}  \,  \mathscr{A}_{z,ijk} \\
\mathscr{A}_{x,ijl}  \,  \mathscr{A}_{y,ijl}  \,  \mathscr{A}_{z,ijl} \\
\mathscr{A}_{x,jkl}  \,  \mathscr{A}_{y,jkl}  \,  \mathscr{A}_{z,jkl} 
\end{pmatrix}
\begin{pmatrix}
B_{x,ijkl} \\
B_{y,ijkl} \\
B_{z,ijkl}
\end{pmatrix}
=
\begin{pmatrix}
\Phi_{ijk}\\
\Phi_{ijl}\\
\Phi_{jkl}\\
\end{pmatrix}
\end{equation}
where the $\mathbf{\mathscr{A}}_{ijk}$ are the outward vector areas of the faces. Note that we are using an integral condition to recover $\mathbf{B}=\nabla\times\mathbf{A}$, so we are not assuming that $\mathbf{A}$ is differentiable, which can be a pitfall for some vector potential numerical schemes for solving the MHD equations.

[4] Finally, the Delaunay tetrahedral magnetic fields are converted to
Voronoi cell magnetic fields $\mathbf{B}_i$ by volume averaging all
the magnetic fields of the tetrahedra that touch cell $i$.

Fig.~\ref{fig:scheme} illustrates the geometrically-averaged
quantities we have defined, and shows a diagram illustrating the steps
required to recover the cell-centred magnetic fields.

We also note that a magnetic field needs to be supplied to the Riemann
solver for the update of the other fluid variables using the finite
volume approach. We found it sufficient to take the extrapolated
cell-centred magnetic fields out to the faces, with the normal
component through the face averaged across the two sides, as done in
\cite{2011MNRAS.418.1392P}.

\subsection{Magnetic vector potential in periodic boundary conditions}\label{sec:A}

Here we describe the  implementation of the magnetic
vector potential approach in a periodic domain. Note that while
$\mathbf{B}$ is periodic, $\mathbf{A}=\nabla\times\mathbf{B}$
need not be. However, in general, the magnetic vector potential may be decomposed into a periodic part and a non-periodic part which corresponds to the mean magnetic field, which is an invariant of ideal MHD.
Thus:
\begin{equation}
\mathbf{A}_i(\mathbf{x},t) = \mathbf{A}_{i,\textrm{mean-field}}(\mathbf{x}) + \mathbf{A}_{i,\textrm{periodic}}(\mathbf{x},t)
\end{equation}

So, to implement periodic boundary conditions, one may keep track of and update $\mathbf{A}_{i,\textrm{periodic}}(\mathbf{x},t)$ and always add to it $\mathbf{A}_{i,\textrm{mean-field}}(\mathbf{x})$, which is static in time and determined by the mean-field.

\section{Numerical Tests}\label{sec:tests}

We test our numerical method on five problems: the classic
Orszag-Tang vortex, the propagation of a circularly polarized Alfv\'en wave, 
driven MHD turbulence, the formation of an idealized magnetized disc, and
a cosmological volume with stellar and black hole feedback.

\subsection{Orszag-Tang}\label{sec:OT}

First, we compare our moving mesh CT method to CT on a static mesh and
moving-mesh Powell cleaning by simulating the classic \cite{1979JFM....90..129O}
vortex, a well-known 2D test problem for
MHD codes that initiates decaying supersonic turbulence. Our set-up of
the problem is described in \cite{2014MNRAS.442...43M}. Here, we
simulate this 2D system inside a 3D box, with all fluid variables
being repeated along the $z$-direction. The initial mesh generating
points are staggered, to yield a non-degenerate Delaunay
tetrahedralization of the mesh (which is not necessary for our method, 
but speeds up initial mesh construction). We simulate the system using
resolutions of $N=16^3$ and $N=64^3$ to evaluate the pre-converged
and resolved behaviours of the methods.

Fig.~\ref{fig:otmain} shows the results of the tests. All the methods
give robust, accurate answers with sufficient resolution. We used an
initially staggered mesh that is not exactly aligned with the
symmetries of the initial conditions, hence giving rise to small
asymmetries in all the simulations. At low/marginally-resolved
resolution, the Powell scheme can be sensitive to the
divergence-cleaning source term, since divergence errors can be the
largest at low resolution, which breaks the symmetry of the problem to
a greater extent.

\subsection{Circularly-polarized Alfv\'en wave}\label{sec:alfven}

We simulate the circularly-polarized Alfv\'en wave of \cite{2000JCoPh.161..605T} in 3D. 
The solution has a known analytic expression, and therefore we can use this test to verify the convergence properties of our scheme. 
The non-linear Alfv\'en wave is chosen to propagate along 
the diagonal of a $2\times 1\times 1$ periodic box of size $3\times 1.5\times 1.5$, as described in \cite{2008ApJS..178..137S}. We use a staggered mesh with resolution $2\times (2N\times N \times N)$.
The problem is initialized with $\rho=1$, $p_{\rm gas}=0.1$, $\Gamma = 5/3$.
In a coordinate frame defined along the diagonal, the wave 
has velocity and magnetic field
$v_\perp=B_\perp=0.1\sin(2\pi x_{\|})$, $v_{\|}=0$, $B_{\|}=1$, 
$v_z=B_z=0.1\cos(2\pi x_{\|})$.
The solution returns to it's original state at time $t=1$.

\begin{figure}
\centering
\includegraphics[width=0.47\textwidth]{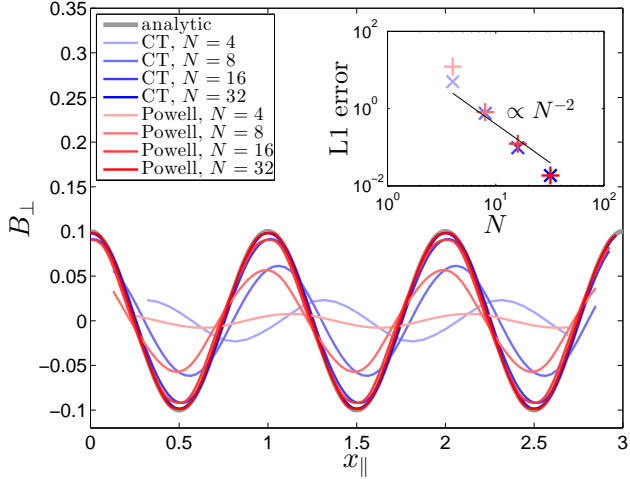}
\caption{Convergence of the non-linear circularly-polarized Alfv\'en wave (3D simulation) with the moving  Powell and CT schemes.}
\label{fig:alfven}
\end{figure}

Fig.~\ref{fig:alfven} shows the convergence of the magnetic field of the Alfv\'en wave at $t=1$ as a function of the resolution $N$. The analysis shows second order behaviour as expected for both the Powell and CT schemes.

\subsection{Turbulent box}\label{sec:turb}

We next simulate isothermal supersonic MHD turbulence using the driving routine of \cite{2010A&A...512A..81F,2010MNRAS.406.1659P}, adapted for the \textsc{Arepo} code in \cite{2012MNRAS.423.2558B}.
The initial conditions are set up in dimensionless units: boxsize $L=1$, sound speed $c_s=1$, initial density $\rho_0=1$, initial magnetic field $\mathbf{B}=(0,0,1)$, and we use $64^3$ resolution elements. We drive the velocity field solenoidally in Fourier-space on large spatial injection scales to establish a turbulent box characterized by a sonic Mach number of $\mathcal{M}_{\rm s}\sim10$ and Alfv\'enic Mach number of $\mathcal{M}_{\rm s}\sim 3$. Saturated turbulence (i.e., with steady time-averaged statistical properties modulo the effects of intermittency) is expected to be established after a few eddy turnover times. 

The results of our simulation are shown in Fig.~\ref{fig:turb}.
We plot a slice of the density field, $x$-direction magnetic field, probability density function (PDF) of the magnetic field strength, and the radially averaged $1D$ power spectrum of the magnetic field at $2$, $10$, and $40$ eddy turnover times. At $2$ eddy turnover times, when the turbulence is transitioning into the non-linear saturated regime, both the CT and Powell scheme show similar features in the density and magnetic field, and the power spectrum (we have used the same random number seed for the turbulent driving). Hence, we know that both codes perform reasonably well in the linear regime of this simulation. The magnetic field power-spectrum is flat on the injection scales and falls over the inertial range. This shape is in good agreement with the magnetic power spectra calculated in \cite{2016arXiv160508662T}, which compares turbulent box simulations using a Cartesian mesh constrained transport scheme (implemented in \textsc{FLASH}) and an SPH method for MHD. 

At later times, the CT and Powell schemes differ. The CT scheme shows approximately time-steady behaviour of the magnetic field PDF and power spectrum, as expected. Additionally, the volume averaged mean magnetic field ($\langle\mathbf{B}\rangle=(0,0,1)$), a conserved quantity in ideal MHD, is well-preserved. Fig.~\ref{fig:turb} lists the mean field values, which are generally preserved to within $1$~per~cent (in fact, they are conserved to machine precision under our scheme on a static mesh, as is the case for the original CT scheme). The Powell scheme, on the other hand, shows poor behaviour in the non-linear regime. The mean magnetic field continues to grow, the magnetic field PDF keeps shifting to the right to higher field strengths, and power is transferred to the largest spatial scale ($k=2\cdot\pi$). The magnetic field have grown by over an order-of-magnitude from its expected value. This is problematic for the simulation: the magnetic field becomes dominant and changes the nature of the turbulence as it transitions from a super-Alfv\'enic to sub-Alfv\'enic regime. 

\begin{figure*}
\centering
\begin{tabular}{cc}
 \large{moving CT}  & \large{moving Powell} \\
\includegraphics[width=0.47\textwidth]{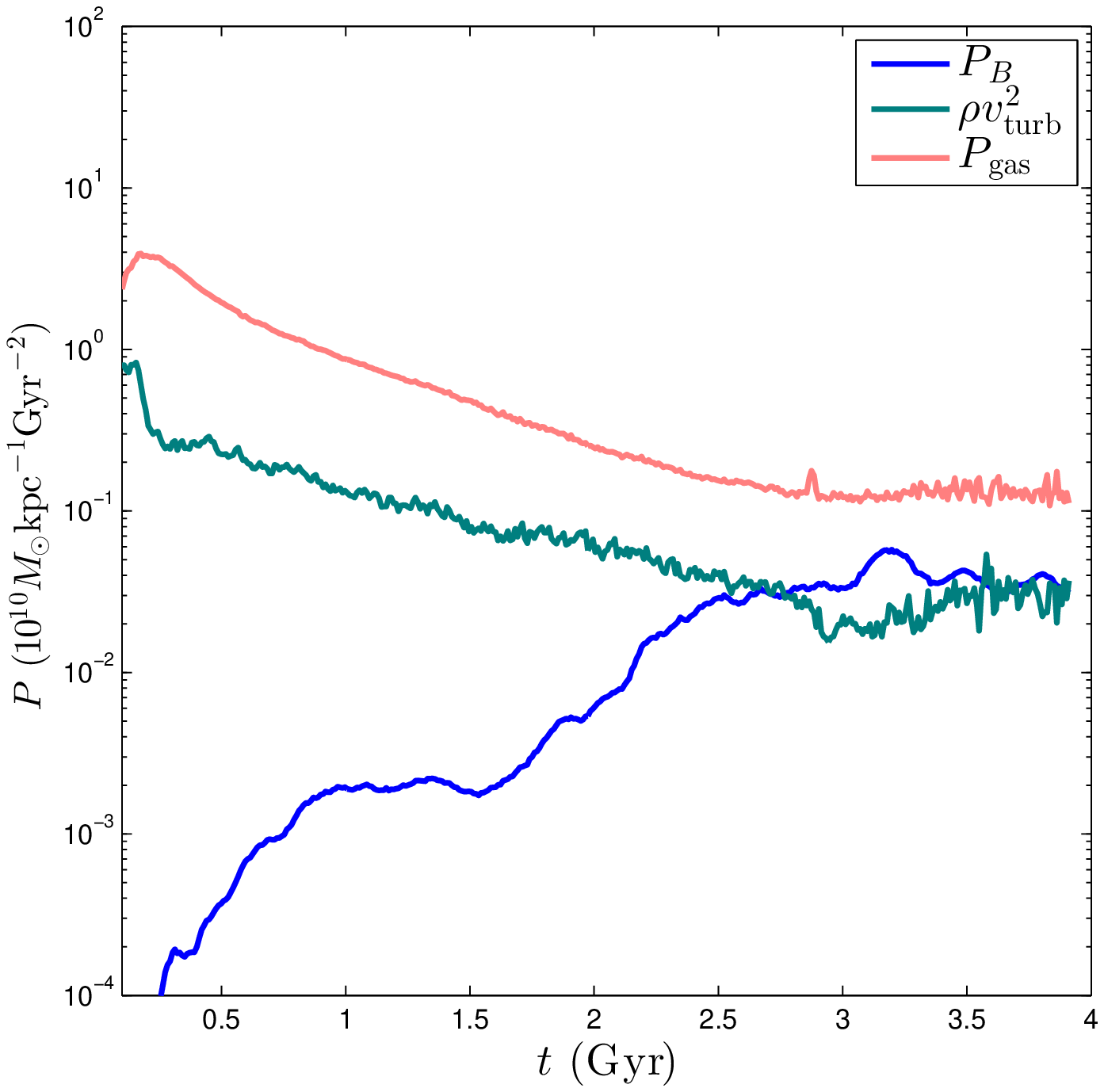} &
\includegraphics[width=0.47\textwidth]{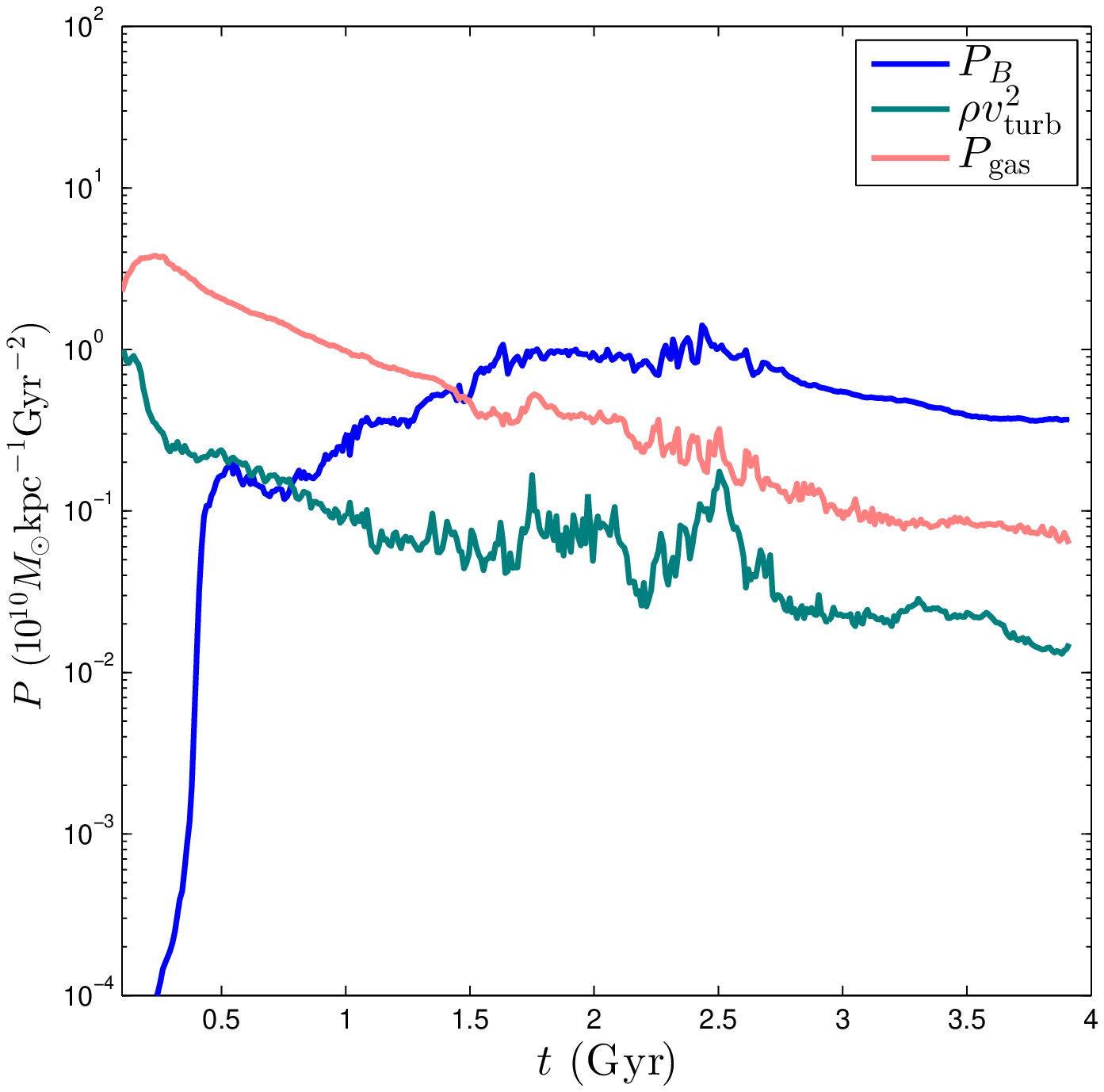}  \\
\end{tabular}
\caption{Comparison of magnetic field saturation in the formation
of a disc simulated with the CT and Powell schemes. The CT method
shows equipartition between magnetic energy density and turbulent
kinetic energy density, whereas the Powell technique saturates 
the field at higher values, exceeding the thermal pressure by about 
a factor of five.}
\label{fig:disc2}
\end{figure*}

\begin{figure*}
\centering
\begin{tabular}{ccc}
 \large{moving CT}  & \large{moving Powell} & \\
\includegraphics[width=0.42\textwidth]{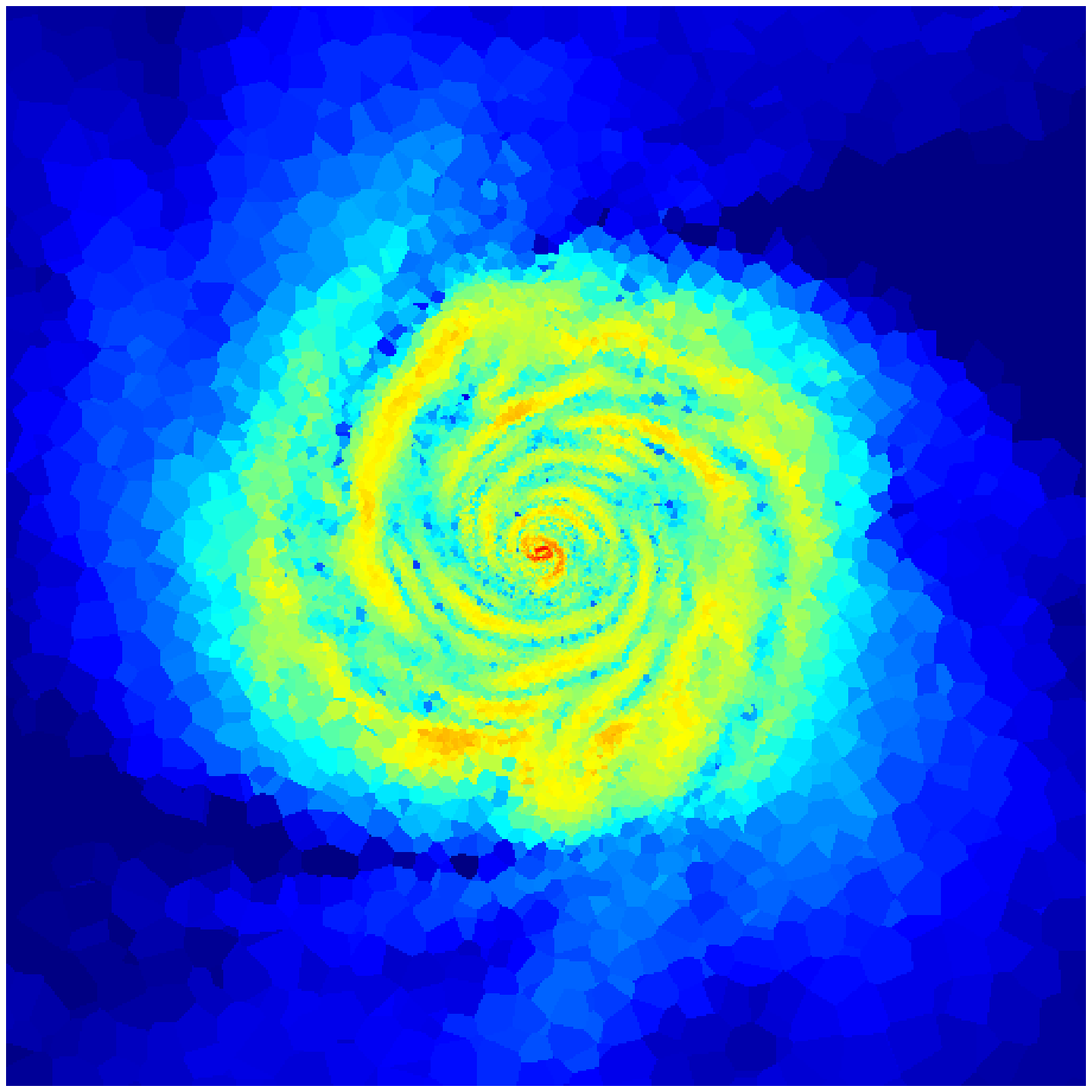} &
\includegraphics[width=0.42\textwidth]{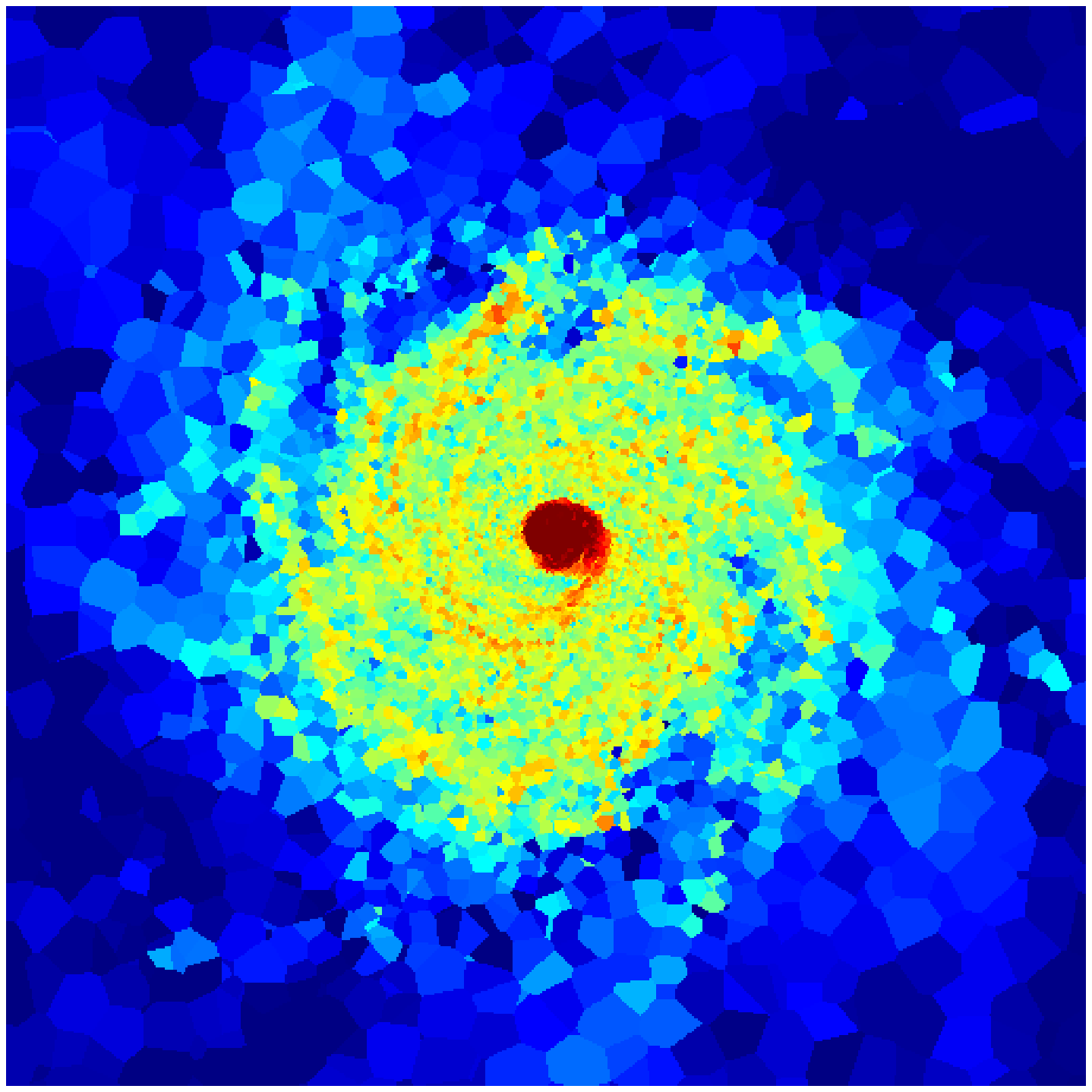} &
\includegraphics[width=0.086\textwidth]{diskcbb.eps} \\
\end{tabular}
\caption{Comparison of the magnetic field strength of the same
disc in Fig.~\ref{fig:disc} at time
$t=0.5~{\rm Gyr}$ in the formation process, simulated using the CT
and Powell schemes. The figure displays a physical size of
$40~{\rm kpc}$. The CT approach exhibits much better preservation of the
topological winding of the magnetic field. The Powell scheme shows 
substantial divergence error noise seen on the cell level while this is absent to machine precision in CT.}
\label{fig:discEarly}
\end{figure*}

\begin{figure*}
\centering
\begin{tabular}{ccc}
 \large{moving CT}  & \large{moving Powell} & \\
\includegraphics[width=0.3\textwidth]{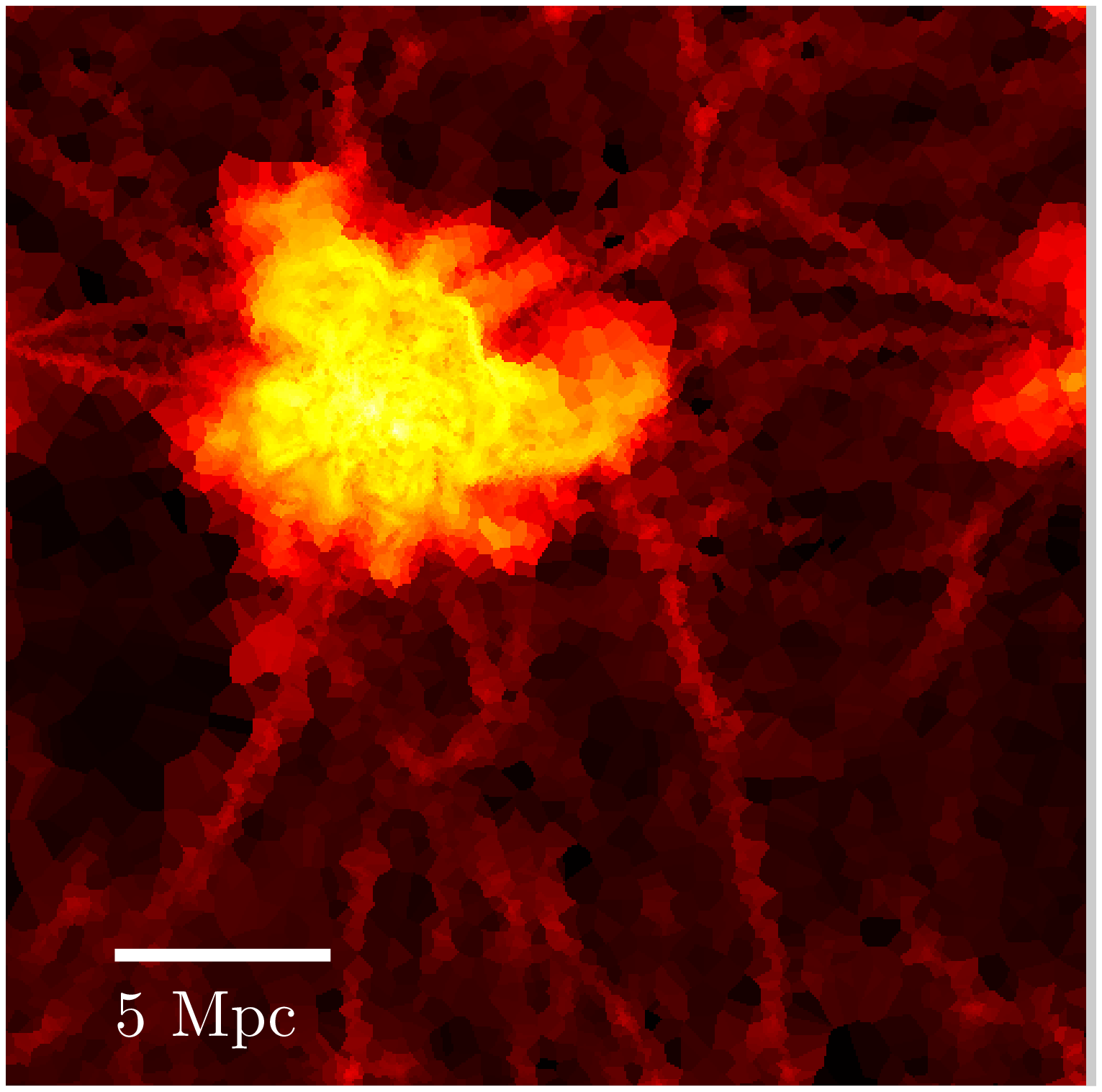} &
\includegraphics[width=0.3\textwidth]{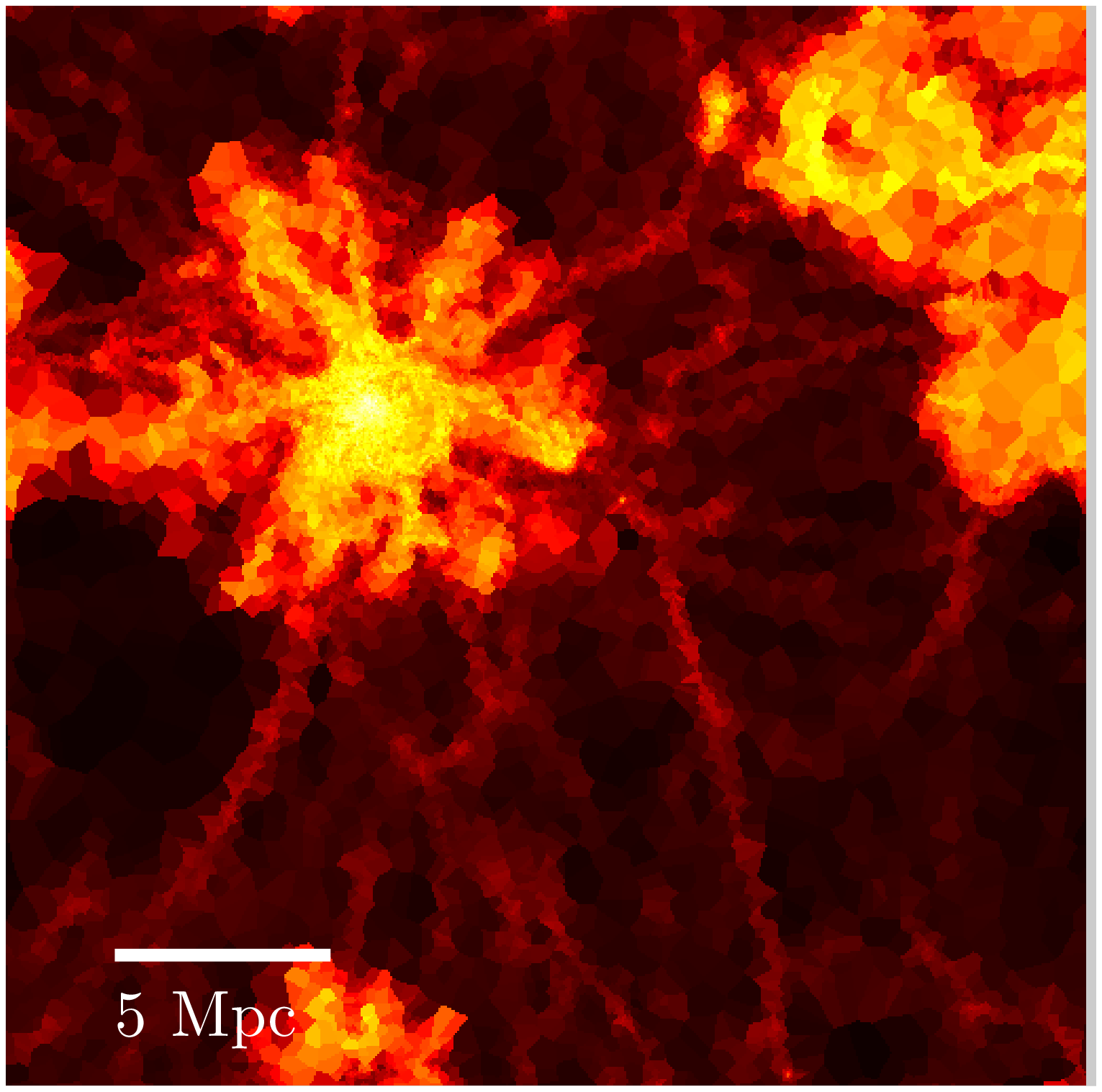} &
\includegraphics[width=0.058\textwidth]{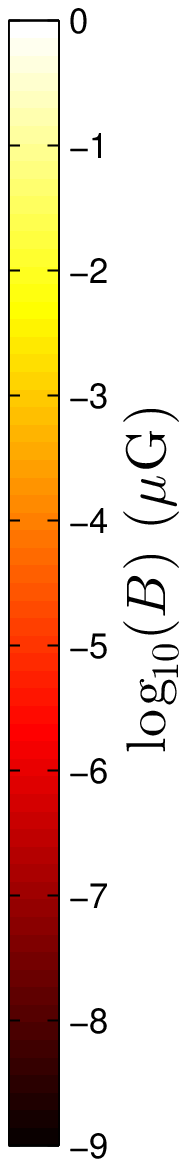} \\
\includegraphics[width=0.3\textwidth]{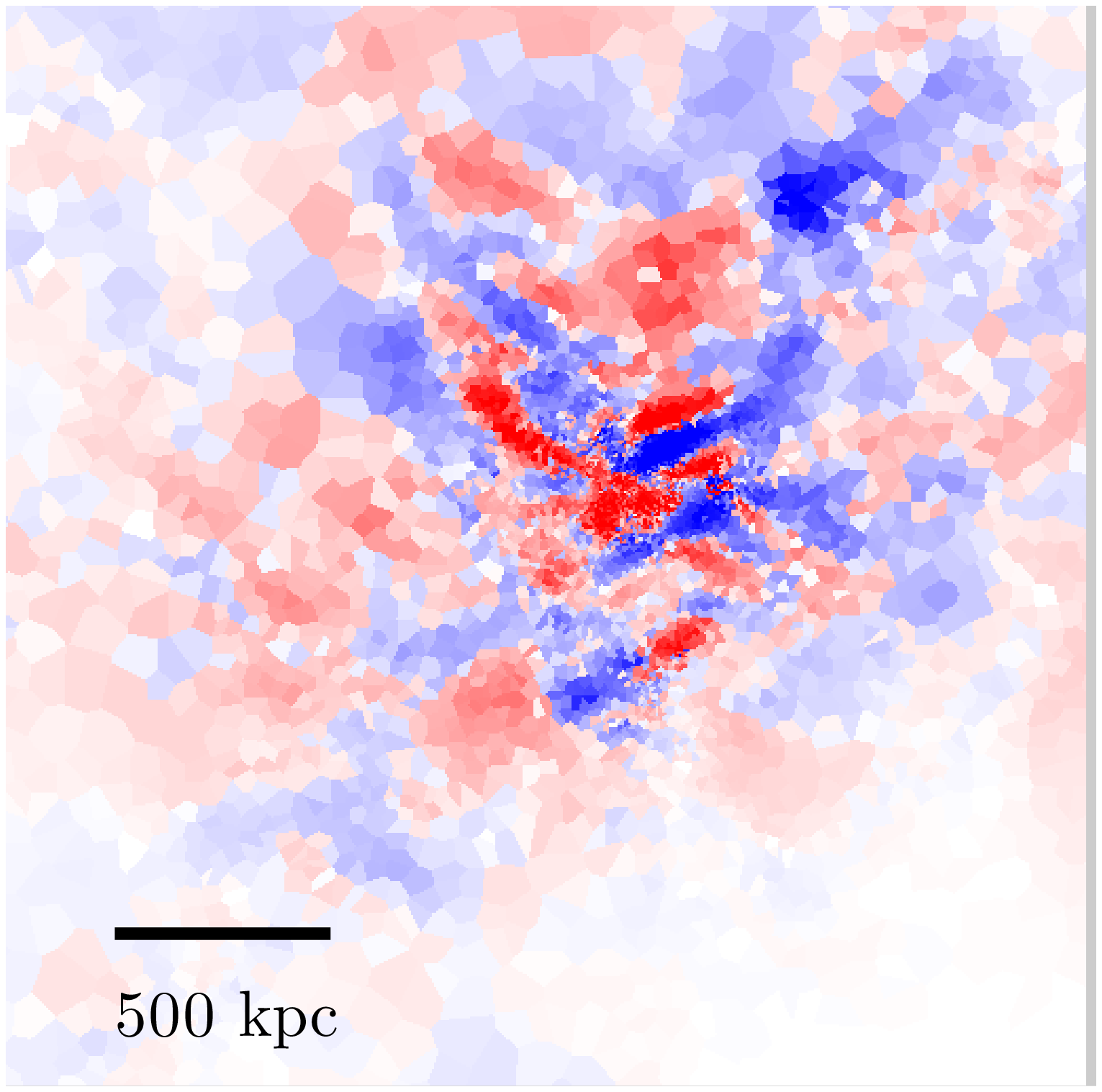} &
\includegraphics[width=0.3\textwidth]{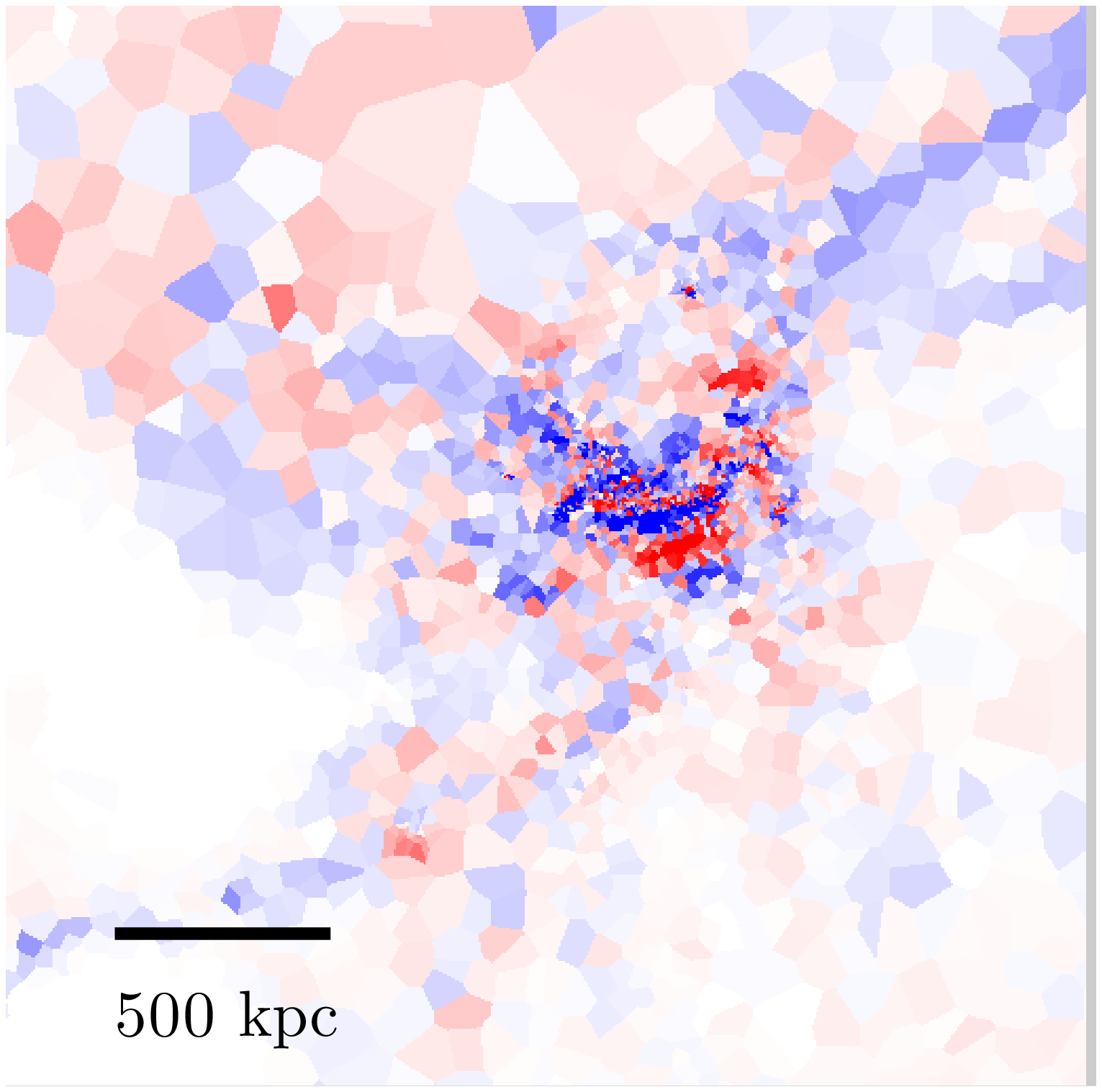} &
\includegraphics[width=0.058\textwidth]{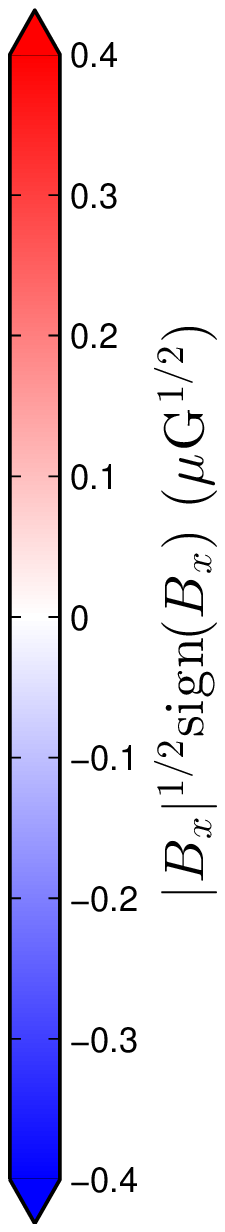} \\
\includegraphics[width=0.4\textwidth]{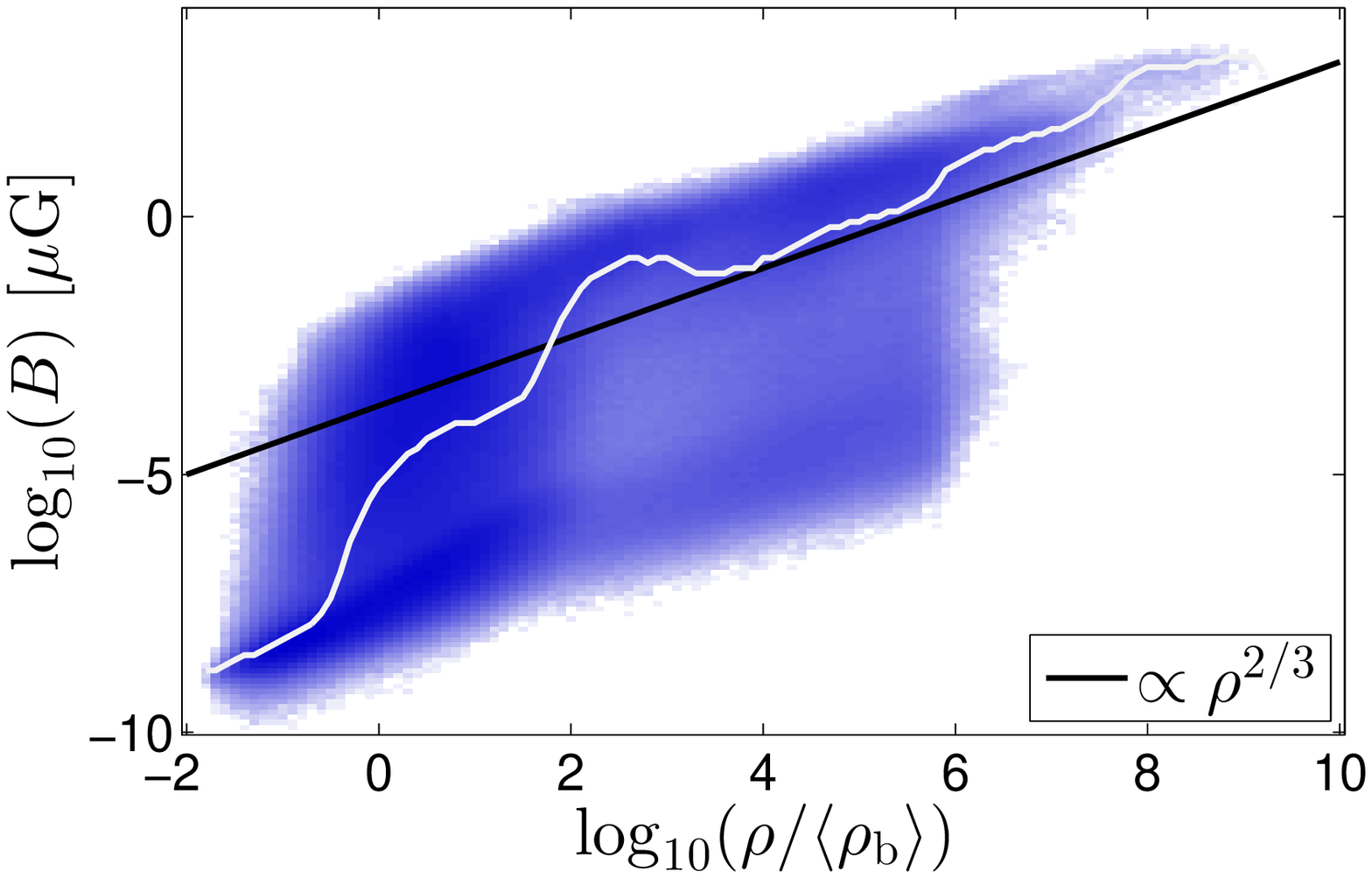} &
\includegraphics[width=0.4\textwidth]{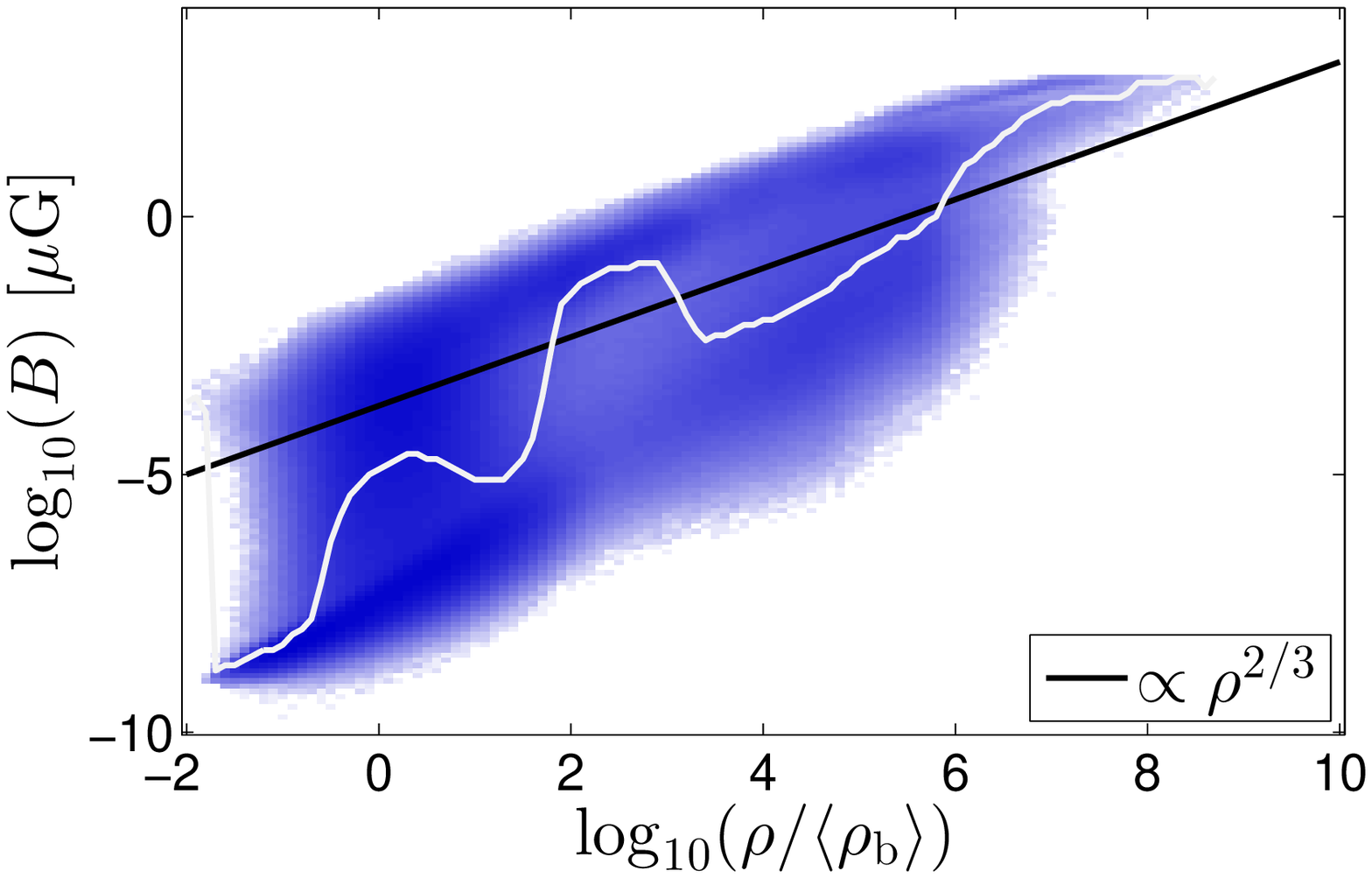} & \\
\end{tabular}
\caption{Magnetic fields (strength and $x$-component, top and
middle panels respectively) and their
phase-space distribution (bottom panels)
at $z=0$, in a $h^{-1}25~{\rm Mpc}$ cosmological box
with full feedback physics as obtained with the CT and Powell
schemes. The magnetic strengths and the phase space diagram are
similar. In the phase-space diagram, the white line
shows the median magnetic field strength at a given density. Under
pure adiabatic compression, $B\propto \rho^{2/3}$ is expected due to
flux-freezing. There is a difference in the magnetic
field topology in the two simulations. The zoom-in on the $B_x$
profile shows that the CT scheme exhibits larger structure and less individual cell-scale noise.}
\label{fig:box25}
\end{figure*}

\subsection{Magnetic Disc}\label{sec:disc}

We simulate the formation of an isolated Milky-Way sized disc galaxy
under idealized conditions, as done previously in
\cite{2013MNRAS.432..176P}, which used the moving Powell method to
evolve the magnetic fields. Here, we compare results from the moving
CT and moving Powell methods. Our simulation set-up uses
$2\cdot10^{5}$ particles (effective mass resolution of $\sim3\times
10^{5}~M_\odot$), and has an initial $3\cdot10^{-10}~{\rm G}$ magnetic
field along the $x$-axis. The total angular momentum vector is
initially along the $z$-axis.

The results comparing our CT method with the Powell cleaning are given
in Figs.~\ref{fig:disc} and \ref{fig:disc2}. We show the density field, magnetic field strength, and $x$ component of
the magnetic field in the disc at $2.5$ Gyr in Fig.~\ref{fig:disc}. We
show the time evolution of the magnetic pressure, thermal pressure,
and turbulent kinetic energy density in Fig.~\ref{fig:disc2}. These
quantities are calculated as volume averaged values inside a
cylinder of radius $15~{\rm kpc}$ and height $1~{\rm kpc}$ centred on the
disc. The turbulent velocity for the calculation of the turbulent
kinetic energy density is calculated by subtracting the mean
rotational velocity (the formed disc has a flat $\sim 200~{\rm
  km}~{\rm s}^{-1}$ rotation curve) from the gas velocities and
computing the root-mean-square.

Using the Powell scheme, \cite{2013MNRAS.432..176P} found that the
magnetic field strength in the disc is quickly amplified, due to
small-scale dynamo action, shearing motions and the central starburst,
and eventually saturates. In this equilibrium state, the magnetic
field pressure equals a few times the thermal pressure. 

Our results with the CT scheme provide a modified picture. The
saturation proceeds more slowly and reaches a lower value. This
asymptotic level is in equipartition with the turbulent kinetic energy
density, in agreement with the theoretical and numerical understanding
of the galactic dynamo and turbulence as well as observations of our
Galaxy \citep{1965PhFl....8.1385K, 1995ApJ...439..779Z,
  1996ARA&A..34..155B,1996ApJ...464..690H}. The topology of the
magnetic field is also quite different. The winding structure of the
magnetic field is preserved to a greater degree with CT. The magnetic
field in the CT method is not dynamically large enough to disrupt the
central parts of the disc and cause strong outflows for our particular
set-up. The winding of the magnetic field is expected, as the
divergence-free condition enforces a topological constraint on the
magnetic field. The ability of a CT code as opposed to Powell
cleaning to maintain topological constraints is shown in
Fig.~\ref{fig:discEarly}, where we plot the magnetic field structure
in the disc at a time of $0.5~{\rm Gyr}$ before much of the magnetic
field amplification due to shear and small-scale dynamo action has
taken place (thus we would expect a smooth profile showing twisting of
the field lines that follows the rotation of the fluid). We see
clearly that the CT algorithm exhibits a winding-structure of the
magnetic field whereas the Powell scheme has considerable noise at the
level of individual mesh cells due to divergence errors. We have quantified the divergence errors relative to the total gas pressure, which describes the dynamical effects of the error. With the CT scheme, the divergence errors are zero
to the level of machine precision, while in the Powell scheme, we have found that the
errors are of order $10$ percent in the central parts of the disc.

\subsection{Cosmological Box}\label{sec:cosmo}

We simulate a $25h^{-1}~{\rm Mpc}$ box cosmological box at $256^3$ 
resolution with a weak initial seed magnetic field of strength 
$10^{-14}~{\rm G}$, using the set-up described in
\cite{2015MNRAS.453.3999M}. The simulation includes the full physics 
(stellar and active-galactic nuclei (AGN) feedback and radiative 
cooling) of the \textsc{Illustris} simulation 
\citep{2013MNRAS.436.3031V,2014Natur.509..177V,2014MNRAS.444.1518V,2015A&C....13...12N}. 
Fig.~\ref{fig:box25} shows the $z=0$ magnetic field strength in a slice 
of the box, the magnetic field $B_x$ component zoomed in on a halo, and 
$\rho$--$B$ phase-space distribution of the gas. The phase-space 
distribution of the magnetic field strengths and the densities 
(normalized by mean baryon density) are fairly similar for the CT and 
Powell methods. The phase-space distribution is expected to follow a 
$B\propto \rho^{2/3}$ under pure adiabatic compression due to the 
flux-freezing condition (solid black line). The magnetic field exceeds 
this value due to amplification by shearing and dynamo action.
\cite{2011ApJ...731...62F} isolate the dynamo action and separating it from the compressional amplification of B, and provide general properties of the dynamo in a compressible gas in \cite{2011PhRvL.107k4504F}.
 Only in halo centres does the magnetic field become strong enough to slightly 
influence the gas dynamics. While magnetic fields are amplified to 
similar strengths using the two numerical methods, their topologies 
differ. The Powell scheme shows more noise at the level of 
individual cells due to divergence errors. Some of the differences in the structure is attributed to the stochasticity of the feedback model.

\subsubsection{Induction equation in comoving coordinates}\label{sec:comov}

We offer remarks on how we chose to implement the induction equation in cosmological comoving coordinates. The global expansion of the universe is characterized by the time-dependent scale factor $a(t)$.
The simulation is evolved on a mesh in comoving variables $\mathbf{x}$, 
and the standard physical fluid variables are also converted to `comoving' variables, which are the quantities being evolved.  
We use the quantities defined in section 2.2 of \cite{2013MNRAS.432..176P}.
The `comoving' magnetic field $\mathbf{B}_{\rm c}$ that our base scheme solves is
related to the physical magnetic field $\mathbf{B}$  by  $\mathbf{B}=\mathbf{B}_{\rm c}a^{-2}$. As \cite{2013MNRAS.432..176P} point out, this choice has the advantage of eliminating source terms in the induction equation.
For our vector potential CT scheme, we define the 
`comoving' vector potential 
$\mathbf{A}_{\rm c}$ by $\mathbf{B}_{\rm c} = \nabla_{\mathbf{x}}\times \mathbf{A}_{\rm c}$. The induction equation is then given by:
\begin{equation}
\frac{\partial \mathbf{A}_{\rm c}}{\partial t} = -\frac{1}{a} \left( -\mathbf{u}\times\mathbf{B}_{\rm c} \right)
\end{equation}
where $\mathbf{u}$ is the peculiar velocity.

\section{Discussion}\label{sec:discussion}

While the Powell and CT schemes both yield accurate results in
idealized test problems (such as the Orszag-Tang vortex and the propagation of a circularly polarized Alfv\'en wave), we find that
the two approaches can lead to different outcomes for astrophysical flows with magnetic field amplification and
turbulence in the non-linear regime. Divergence cleaning techniques have been shown to produce
spurious structures and magnetic energy fluctuations on small spatial
scales due to the non-locality of the cleaning step
\citep{2000JCoPh.161..605T,2004ApJ...602.1079B}, whereas CT schemes
produce robust results.

In moving mesh simulations of turbulent and astrophysical flows, the Powell and CT
schemes give a similar general picture, but exhibit some quantitative
differences. \cite{2015ApJ...806L...1Z} discusses that Powell cleaning
gives faster field growth and larger saturation at lower resolutions
in general, where the divergence errors are larger. This is consistent
with our results for the magnetized disc. The CT method reaches the
natural equipartition between the magnetic energy density and the
turbulent kinetic energy density in the disc, whereas the Powell
scheme overshoots this value and the magnetic pressure dominates the
gas pressure by a factor of five. In the original Powell magnetic disc
simulations of \cite{2013MNRAS.432..176P}, the authors find even
larger ($>10\times$) overshoots of the magnetic pressure over gas
pressure in their lowest resolution simulation ($12500$ particles; see
their Figure~7). The CT scheme can be considered particularly more
robust at low/marginally-resolved resolutions, which will always be
the situation in practice for some structures in large-scale
cosmological simulations.

This difference between CT and Powell in the growth rate and saturation 
of the magnetic field in magnetized disc galaxies is observed across different 
codes in the literature as well. 
Adaptive mesh refinement (AMR) with the divergence-preserving CT scheme \citep{2009ApJ...696...96W,2010A&A...523A..72D,2016MNRAS.457.1722R} find slow 
growth rates compared to the fast growth rates (e-folding time $\sim10$~Myr) of 
divergence cleaning schemes implemented in SPH \citep{2012MNRAS.422.2152B}
 or moving mesh \citep{2013MNRAS.432..176P}. The difference we observe between 
 our CT and Powell simulations suggests that divergence cleaning, and not 
 the Eulerian vs Lagrangian natures of the codes, is the culprit. 
 \cite{2016MNRAS.457.1722R} point out that in simulations without strong stellar 
 feedback (such as ours), the flows are relatively smooth and two-dimensional, a regime in which 
 strong dynamo amplification is difficult.

In the case of the cosmological box simulations, the two schemes give 
similar statistics for the overall magnetic field strengths in the density - magnetic field phase space. However, 
the CT algorithm produces a smoother magnetic field, without the 
magnetic field unphysically flipping at the single-cell level. This 
property will improve Faraday rotation measure estimates and cosmic ray 
propagation \citep{2016arXiv160407399P,2016arXiv160500643P} in the \textsc{Arepo} code.

The difference between the moving mesh CT and Powell schemes can perhaps 
be best understood with the idealized turbulence test, where the behaviour of the true solution is known theoretically.
The Powell scheme, with its non-conservative formulation (divergence correcting source-terms), 
artificially makes the magnetic field grow in this particular test problem and quickly transfers magnetic energy to the largest scales. The mean magnetic field (which should be a conserved quantity in ideal MHD as the equations for the evolution of the magnetic field can be recast into standard conservative hyperbolic form) grows by over an order of magnitude and information about the initial mean field direction is also lost. Interestingly, the disc simulation shows similar difference between the CT and Powell schemes as these turbulence tests; namely the Powell scheme grows the magnetic field beyond the turbulent kinetic energy and  most of the magnetic power is on the largest scales (size of the disc).

Finally, we also note that our unstructured (cell-centred) CT formalism may find 
useful application in static mesh codes, and may potentially be used to 
simplify some of the technical details of implementing CT in a code 
with AMR
\citep{2001JCoPh.174..614B,2006A&A...457..371F,2012ApJS..198....7M} as 
well.

A moving mesh unstaggered CT approach has some practical advantages 
over an AMR code with staggered CT. The unstaggered representation 
allows for straightforward coupling with refinement and power of two 
hierarchical time stepping on a moving Voronoi mesh. There is no 
difficulty introduced as in the fine and coarse level mismatch in AMR 
codes, where specific care (such as restriction, prolongation, and 
reflux-curl operations \citealt{2011ApJS..195....5M}) has to be taken 
to prevent a loss of order of accuracy and breaking of the 
divergence-free condition at interface regions. In addition, the moving 
mesh approach provides the usual advantages of automatic adaptivity and 
reduced advection errors.

Summarizing, we have developed an efficient, accurate, unstructured CT scheme for 
solving the MHD equations in 3D on a moving mesh. The CT formulation 
allows for the maintenance of the divergence-free condition on the 
magnetic field to machine precision, which leads to numerical 
stability, more accurate numerical solutions, and good preservation of 
the topological properties of the magnetic field. The numerical experiments 
we considered demonstrate the advantages of CT schemes over cleaning schemes, 
namely preventing artificial magnetic field growth due to the source terms in cleaning 
schemes. The new CT method, 
implemented in the moving mesh code \textsc{Arepo}, will allow for 
powerful MHD simulations in the upcoming future, making use of the advantages of 
an adaptive, quasi-Lagrangian treatment of the fluid equations.

\section*{Acknowledgements} This material is based upon work supported 
by the National Science Foundation Graduate Research Fellowship under 
grant no. DGE-1144152. PM is supported in part by the NASA Earth and 
Space Science Fellowship.  LH acknowledges support from NASA grant 
NNX12AC67G and NSF grant AST-1312095. RP and VS acknowledge support through the European Research Council under ERC-StG grant EXAGAL-308037, and thank the Klaus Tschira Foundation. VS also acknowledges subproject EXAMAG of the Priority Programme 1648 'Software for Exascale Computing' of the German Science Foundation. The computations in this paper 
were run on the Odyssey cluster supported by the FAS Division of 
Science, Research Computing Group at Harvard University.

\bibliography{mybib}{}

\appendix
\section{Other gauge choices}
\label{appendix}

The evolution of the magnetic field is invariant under a choice of gauge $\psi$ for the vector potential:
\begin{equation}
\frac{\partial \mathbf{A}}{\partial t} = -\mathbf{E} - \nabla\psi
\end{equation}
In the present work, we chose to evolve the vector potential under the \textit{Weyl} gauge ($\psi=0$), as it offers a nice correspondence between CT and vector potential methods because one is adding just the same EMF terms to update the solution in both cases. However, it may be of some interest to explore other gauge choices in future work, as they can be more suitable to the treatment of particular problems/physical processes. We briefly discuss alternate choices here. 

One may choose the \textit{Coulomb} gauge, $\nabla\cdot\mathbf{A}=0$. This gauge choice would require using either projection methods or cleaning methods on $\mathbf{A}$ to keep it divergence free. This choice of gauge would simplify the extension of the ideal MHD code to resistive MHD, as the induction equation for resistive MHD is given by:
\begin{equation}
\frac{\partial \mathbf{A}}{\partial t} = \mathbf{v}\times\mathbf{B} - \eta\nabla\times(\nabla\times\mathbf{A} )
\end{equation}
where the resistive term expands as: $\eta \nabla(\nabla\cdot\mathbf{A})-\eta\nabla^2\mathbf{A}$, which is just a diffusion term if $\nabla\cdot\mathbf{A}=0$.

Another choice would be to choose a \textit{helicity-like} gauge: $\psi=\mathbf{v}\cdot\mathbf{A}$. This choice of gauge makes the magnetic vector potential induction equation Galilean-invariant:
\begin{equation}
\frac{d\mathbf{A}}{dt} = -\mathbf{A}\times\nabla\times\mathbf{v} - (\mathbf{A}\cdot\nabla)\mathbf{v} = -\boldsymbol{J}_{\mathbf{v}}^{\rm T}\mathbf{A}
\end{equation}
where $\boldsymbol{J}_{\mathbf{v}}$ denotes the Jacobian of $\mathbf{v}$. The downside of this gauge is that the velocity field may not be differentiable, so special care would have to be taken for the treatment of the Jacobian term in the presence of shocks.

A final interesting choice would be to modify the above helicity-like gauge to use the mesh motion velocity $\mathbf{w}$ instead:
$\psi=\mathbf{w}\cdot\mathbf{A}$, in which case the induction equation becomes
\begin{equation}
\frac{d\mathbf{A}}{dt} = (\mathbf{v}-\mathbf{w})\times\mathbf{B} -\boldsymbol{J}_{\mathbf{w}}^{\rm T}\mathbf{A}
\end{equation}
The first term $(\mathbf{v}-\mathbf{w})\times\mathbf{B}$ is just the electric field in the frame of the moving mesh, which is $0$ if the mesh cell moves exactly with the fluid. This choice of gauge reduces to the Weyl gauge in the case of a static mesh. Such a gauge would require smoothed mesh motion (e.g. \citealt{2015MNRAS.449.2718D}) to treat the Jacobian term accurately and avoid mesh noise.

%###########################################################################################################
\bsp
\label{lastpage}
\end{document}